\documentclass[12pt,a4paper,final,twocolumn]{iopart}
\usepackage{graphicx}
\usepackage{iopams}                
\usepackage{longtable}
\newcommand{\apjs}{{\it Astrophys. J. Suppl.} }
\newcommand{\apj}{{\it Astrophys. J.} }
\newcommand{\aas}{{\it Astron. Astrophys. Suppl.} }
\newcommand{\asa}{{\it Astron. Astrophys.} }
\newcommand{\pra}{{\it Phys. Rev. A} }
\newcommand{\cpc}{{\it Comput. Phys. Commun.} }
\def\rma{{\rm a}}
\def\rmb{{\rm b}}
\def\rmc{{\rm c}}

\def\rms{{\rm s}}
\def\rmp{{\rm p}}
\def\rmd{{\rm d}}
\def\rmS{{\rm S}}
\def\rmP{{\rm P}}
\def\rmD{{\rm D}}
\def\rmF{{\rm F}}
\def\rmG{{\rm G}}

\begin{document}

\title[Inner-shell excitation of Fe$^{15+}$]
{${\bi R}$-matrix inner-shell electron-impact excitation of
Fe$^{15+}$ including Auger-plus-radiation damping}

\author{G Y Liang\footnote{E-mail:
guiyun.liang@strath.ac.uk}, A D Whiteford and N R Badnell}

\address{Department of Physics, University of Strathclyde,
Glasgow, G4 0NG, UK}

\begin{abstract} We present results for the inner-shell electron-impact
excitation of Fe$^{15+}$ using the intermediate-coupling frame
transformation {\it R}-matrix approach in which Auger-plus-radiation damping
has been included.
The target and close-coupling expansions are both taken to be the 134
levels belonging to the configurations ${\rm 2s^22p^63}l$, ${\rm
2s^22p^53s3}l$, ${\rm 2s^22p^53p^2}$ and ${\rm 2s^22p^53p3d}$. The
comparison of Maxwell-averaged effective collision strengths
with and without damping shows that the
damping reduction is about 30-40\% for many transitions at
low temperatures, but up to 80\% for a few transitions.
As a consequence,  the results of previous Dirac $R$-matrix
calculations (Aggarwal and Keenan, 2008) overestimate the
effective collision strengths due to their omission of
Auger-plus-radiation damping.
\end{abstract}
\submitto{\jpb}
\maketitle
\pacs{34.80 Kw}

\section{Introduction}
Radiation from Fe$^{15+}$ occupies a considerable fraction of the
EUV and X-ray radiation spectrum of (the astrophysically abundant
element) iron. Its temperature of peak fractional abundance is at
$\approx 2.5\times10^6$~K in collision dominated plasmas~(Bryans
\etal 2006) and a few times $10^4$~K in photoionized plasmas
(Kallman and Bautista 2001). Observations of inner-shell
excitation lines such as ${\rm 2p^63s}$--${\rm 2p^5}3l3l'$ in the
solar corona have led to extensive investigations of inner-shell
excitation data for this ion (see, e.g., Dere \etal 2001).

Earlier calculations adopted the distorted-wave (DW) approximation.
For example, Cornille \etal (1994) reported excitation data
amongst the lowest 44 levels belonging to the ${\rm 2p^63s}$ and
${\rm 2p^53s3}l$ ($l$=s, p and d) configurations. This data was
adopted by Phillips \etal (1997) to analyze the contribution of
satellite lines to line-ratios, arising from inner-shell
excitations, which are useful in solar diagnostic
applications. Resonant-excitation plays an important role in
electron-ion collision processes, enhancing the effective
collision strength ($\Upsilon$), especially for forbidden
transition lines. These lines are usually
density- and temperature-sensitive and so have potential
diagnostic applications. Bautista~(2000) performed a standard
$R$-matrix (Berrington \etal 1995) calculation for inner-shell
excitation which included the 134-levels belonging to the ${\rm
2s^22p^63}l$, ${\rm 2s^22p^53s3}l$, ${\rm 2s^22p^53p^2}$ and ${\rm
2s^22p^53p3d}$ ($l$=s, p and d) configurations (the same configurations
considered in the present work). The enhancement of
Maxwell-averaged effective collision strengths ($\Upsilon$) by resonances
in the ordinary collision strengths ($\Omega$) was found to be up
to three orders of magnitude
at low temperatures, for some transitions. In Bautista's calculation,
relativistic effects were included by using term-coupling
coefficients (TCC) via the {\sc jajom} code \footnote{The TCCs were obtained
from the $R$-matrix {\sc recupd} code.}. This changed the background
collision strengths by up to an order of magnitude when compared to
the results of his $LS$-coupling calculations, in which the algebraic
splitting of scattering matrices was used to obtain the
fine-structure data. Recently, Aggarwal \& Keenan~(2008)
calculated inner-shell excitation data using the same
134-level target configurations with the fully-relativistic Dirac
atomic {\it R}-matrix code ({\sc darc}) of Norrington and Grant~(1987).
Detailed comparisons with the excitation data of Bautista (2000)
were made, and they pointed out deficiencies in the data of
Bautista due to the methodology used by {\sc jajom}.

In a detailed study of Fe$^{14+}$, Berrington \etal (2005) found
that the Breit--Pauli $R$-matrix effective collision strengths
agreed with the {\sc darc} calculations to within 6\%. For complex
species, the number of (closely spaced) levels that must be
included in the close-coupling (CC) expansion is very large, which
makes the calculation computationally demanding. An alternative
approach to a full Breit--Pauli $R$-matrix calculation is to
perform an $R$-matrix calculation in $LS$-coupling and then, on
making use of multi-channel quantum defect theory (MQDT),
transform the resulting `unphysical' $K$- or $S$-matrices to
intermediate coupling. This eliminates at root the deficiency of
{\sc jajom}, viz. only transforming the open--open part of the
physical $K$-matrix, since all channels are treated as being
`open' in MQDT. This is the intermediate coupling frame
transformation (ICFT) method. In studying of the ICFT $R$-matrix
electron excitation of Fe$^{14+}$ and Ni$^{4+}$, Griffin \etal
(1998) and Badnell \& Griffin (1999) found that the ICFT results
agreed closely with those determined from the full Breit-Pauli
$R$-matrix calculation. Another advantage of the ICFT method is
the saving of computational time, which  makes meaningful
iso-electronic sequence calculations a reality within the
$R$-matrix framework (Witthoeft \etal 2007).

Resonances superimposed upon the background cross section enhance
the effective collision strengths for electron-impact excitation,
especially at lower temperatures and/or for weaker transitions.
However, some resonant states may decay by an Auger process or
fluorescence radiation and so are lost to the transition under study.
Such loss mechanisms can be represented by a complex optical potential.
Robicheaux \etal (1995) provided a detailed description of radiation damping via 
such a potential within the $R$-matrix method. Subsequently,
Gorczyca \etal (1995) showed the effect of radiation damping on the electron-impact
cross section of Ti$^{20+}$ while Gorczyca \& Badnell (1996)
demonstrated its even greater importance for photorecombination. Gorczyca
and Robicheaux (1999) extended the optical potential approach so
as to include Auger damping. 
Whiteford \etal (2002) demonstrated the Auger damping effect on the effective
collision strengths of inner-shell transitions in Li-like Ar$^{15+}$ and
Fe$^{23+}$, and showed significant reductions in effective collision
strengths at low temperatures ($\sim$30\% for the ${\rm 1s^22s}$ ${\rm
^2S_{1/2}}$ -- ${\rm 1s2s^2}$ ${\rm ^2S_{1/2}}$ transition of
Fe$^{23+}$). Correspondingly, this has an influence on the
spectroscopic diagnostic and modelling of plasmas, especially photoionized
plasmas which typically have a much lower electron temperature. Furthermore,
Bautista \etal (2004) demonstrated the smearing of the photoabsorption
K-edge by such damping, primarily Auger, for Fe$^{16+}$ through Fe$^{22+}$.

In the present work we study the inner-shell electron-impact excitation
of Fe$^{15+}$, via the $R$-matrix ICFT approach, using the same CC
and CI expansions as in the work of Aggarwal \& Keenan (2008) but
now include Auger-plus-radiation damping. This work is a part of
on-going collaborative work --- the UK atomic processes for astrophysical
plasmas (APAP) network\footnote{http://amdpp.phys.strath.ac.uk/UK\_APAP},
a broadening of scope of the original UK RmaX network. In
section 2 we present details of our structure calculation and make
comparisons with other data available in the literature.
Our calculations for the scattering problem are detailed
in section 3. The results, and their comparison with those of others, are
discussed in section 4.  We conclude with section 5.

\section{Structure}
We included the following configurations: ${\rm 2s^22p^63}l$,
${\rm 2s^22p^53s3}l$, ${\rm 2s^22p^53p^2}$ and ${\rm
2s^22p^53p3d}$. The orbital basis functions (1s -- 3d) were
obtained from {\sc autostructure} (Badnell 1986) using the
Thomas-Femi-Dirac-Amaldi model potential (Eissner \etal 1974). The radial scaling
parameters were obtained by a two-step procedure of energy
minimization. In the first step, the average energy of all 59 terms
was minimized by allowing all scaling parameters (one for each $nl$
orbital) to change. We then fixed the resulting radial scaling parameter
of the 1s orbital ($\lambda_{1\rms}$=1.41958). Finally, we minimized the average 
energy sum of all 134 levels, obtained from an intermediate coupling calculation,
so as to determine the remaining scaling parameters. 
The resultant values are $\lambda_{2\rms}$=1.30324,
$\lambda_{2\rmp}$=1.14032, $\lambda_{3\rms}$=1.23627,
$\lambda_{3\rmp}$=1.13555 and $\lambda_{3\rmd}$=1.00615.
(The mass-velocity plus Darwin contribution from the 1s orbital
is too large for the minimization procedure to converge if the
1s scaling parameter is varied in intermediate coupling --- the
energy functional has no minimum.)

We compare our energies with the values available from the NIST
database v3.0~\footnote{http://physics.nist.gov/PhysRefData/ASD/levels\_form.html}
and the {\sc grasp} calculation of Aggarwal \& Keenan (2007) (hereafter
referred to as AK07). Their (AK07) calculations of structure used the same 
configurations as herein. The subsequent electron collision scattering 
calculations of Aggarwal \& Keenan 2008 (hereafter, AK08) also used a
structure determined by AK07. Excellent agreement (within 0.1\%) is obtained when 
compared with the results of the AK07 {\sc grasp} calculation that
omitted Breit and QED effects. 
The AK07 data is systematically higher than our results by less than 0.1~Ryd
for the doubly-excited levels.
The agreement with the NIST data is to
within 0.5\%, except for the ${\rm 2s^22p^53s3d~^2D_{5/2}}$, ${\rm
^2F_{7/2}, ^2F_{5/2}}$ and ${\rm ^2P_{3/2}}$ levels. 
The difference is within 0.2\% for these levels. Note, although AK07 obtained better agreement
with the NIST data when Breit and QED effects were included, they are not present
within {\sc darc} (nor two-body fine-structure within Breit--Pauli $R$-matrix)
and so such a structure cannot be used in a scattering calculation.

\begin{figure*}[t]
\hspace{-2.5cm}\includegraphics[angle=0,width=11.cm]{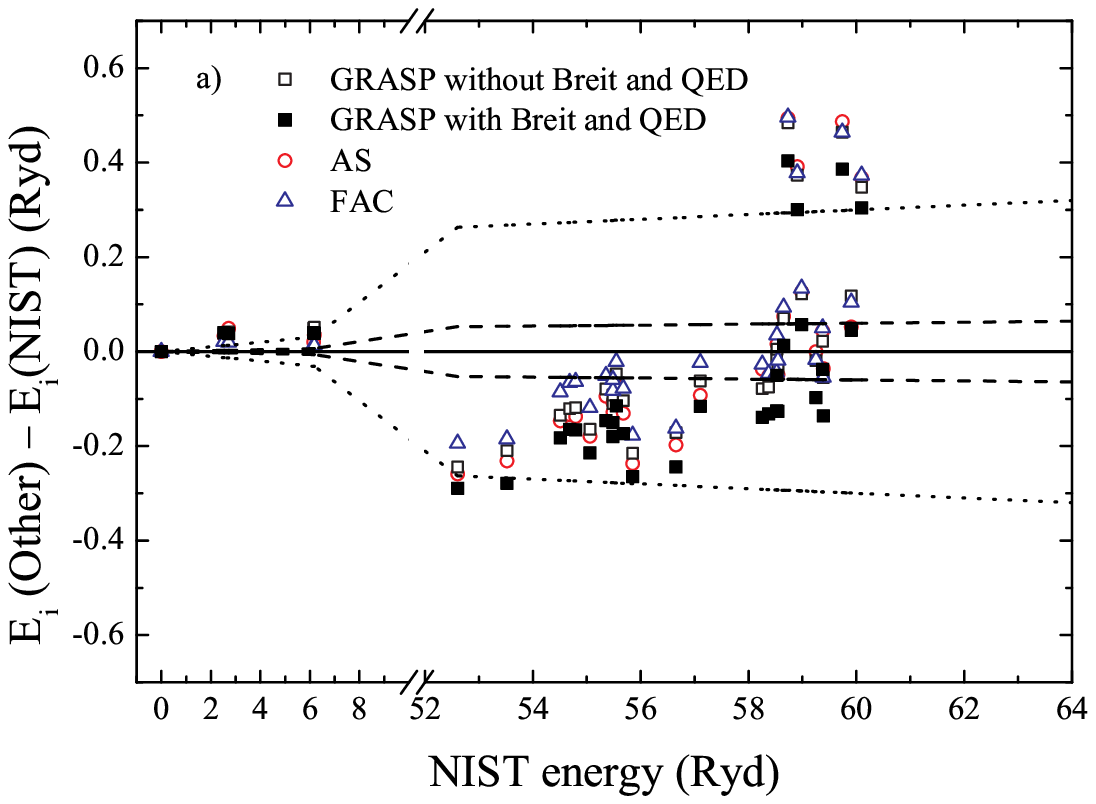}
\hspace{-1.0cm}
\includegraphics[angle=0,width=11.cm]{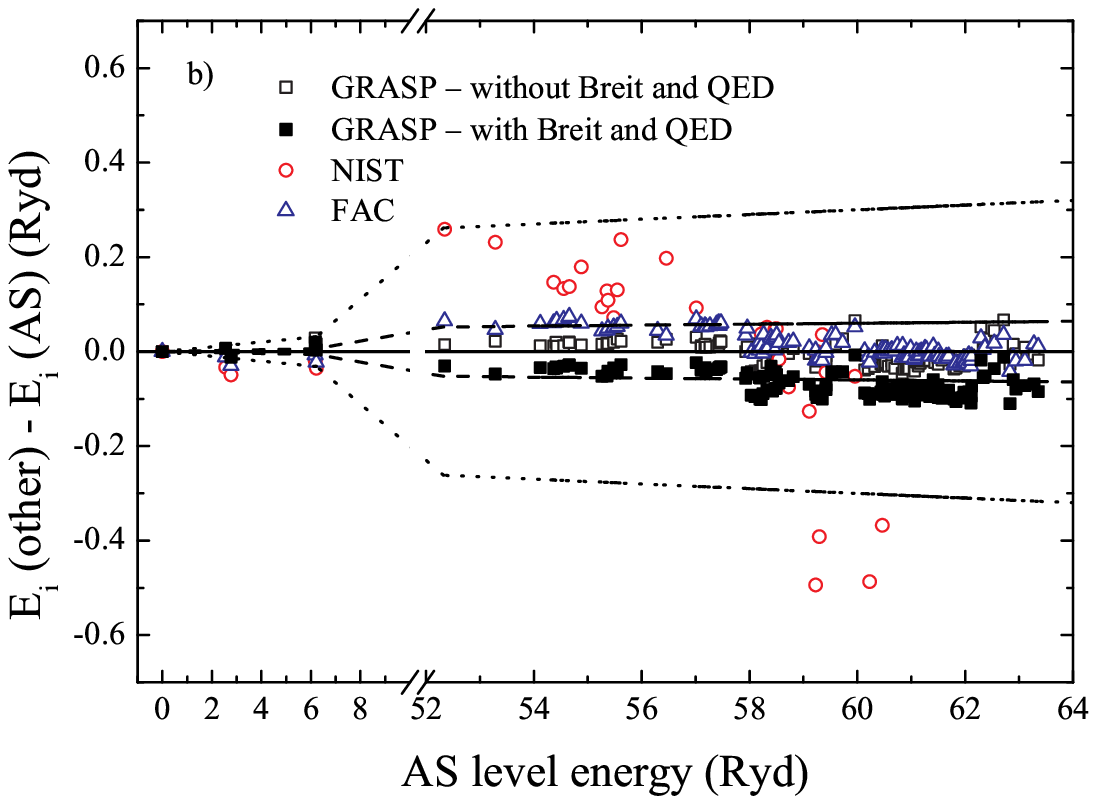}
\caption{Comparisons of energy levels from the present {\sc
autostructure} (AS), {\sc grasp} (Aggarwal \& Keenan, 2007) and
{\sc fac} (present work) calculations. a): differences relative
to the available experimental values (NIST database) versus the
NIST data. Opened and filled square symbols correspond to the {\sc
grasp} results, without and with the inclusion of Breit and QED
effects, respectively (Aggarwal \& Keenan 2007); triangle symbols
indicate the present {\sc fac} results; opened circles denote
the present AS results. b): The differences relative to the
present AS results versus the present AS results. Symbols as before.
The dashed and dotted lines correspond to agreement within 0.1\%
and 0.5\%, respectively. [{\it Colour online}]} \label{fig_level}
\end{figure*}

Due to the strong configuration interaction and level mixing, as
illustrated in table 1 of AK07, level orderings for comparisons
are not same in different calculations. Here, we match the level
assignment according to configuration, total angular momentum and
then energy ordering. Fortunately, only a few level assignments
are inconsistent in the two different calculations, which
facilitates our later comparisons for radiative decay rates
($A$-coefficients) and collision strengths. However, for some
levels, their different assignments result in large discrepancies
(up to 0.84~Ryd) with the NIST values in the AK07 work, such as
for levels 11 and 14. Similar disturbed level ordering appears for
levels 19/20, 26/27, 62/63 etc. Our new assignment eliminates the
mis-order compared to the NIST values, as shown in
table~\ref{energy}.

We also performed a structure calculation with the flexible atomic
code ({\sc fac}) of Gu (2003), which shows slightly better
agreement with our {\sc autostructure} results than those from
{\sc grasp}. Both are systematically higher than {\sc grasp}'s,
{\sc fac} more so. The results of the three different calculations
are compiled in table~\ref{energy}, along with NIST data.

\begin{figure*}[t]
\hspace{-1.cm}\includegraphics[angle=0,width=9.5cm]{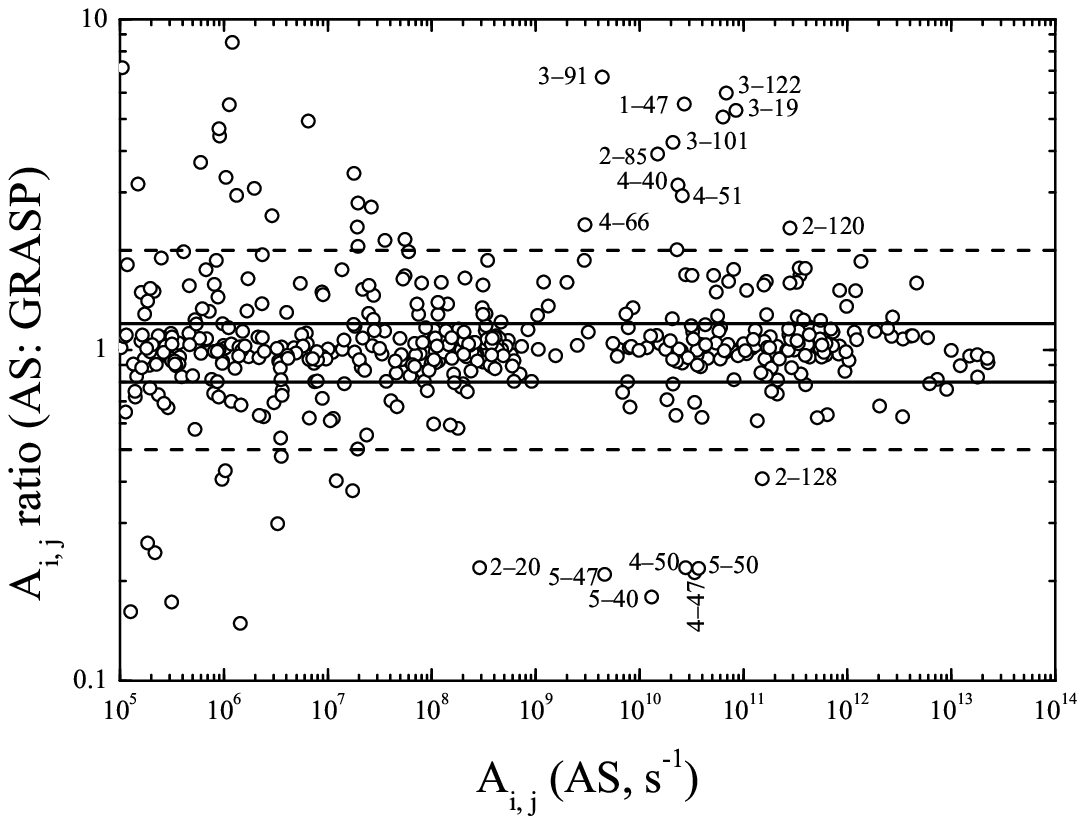}
\includegraphics[angle=0,width=9.5cm]{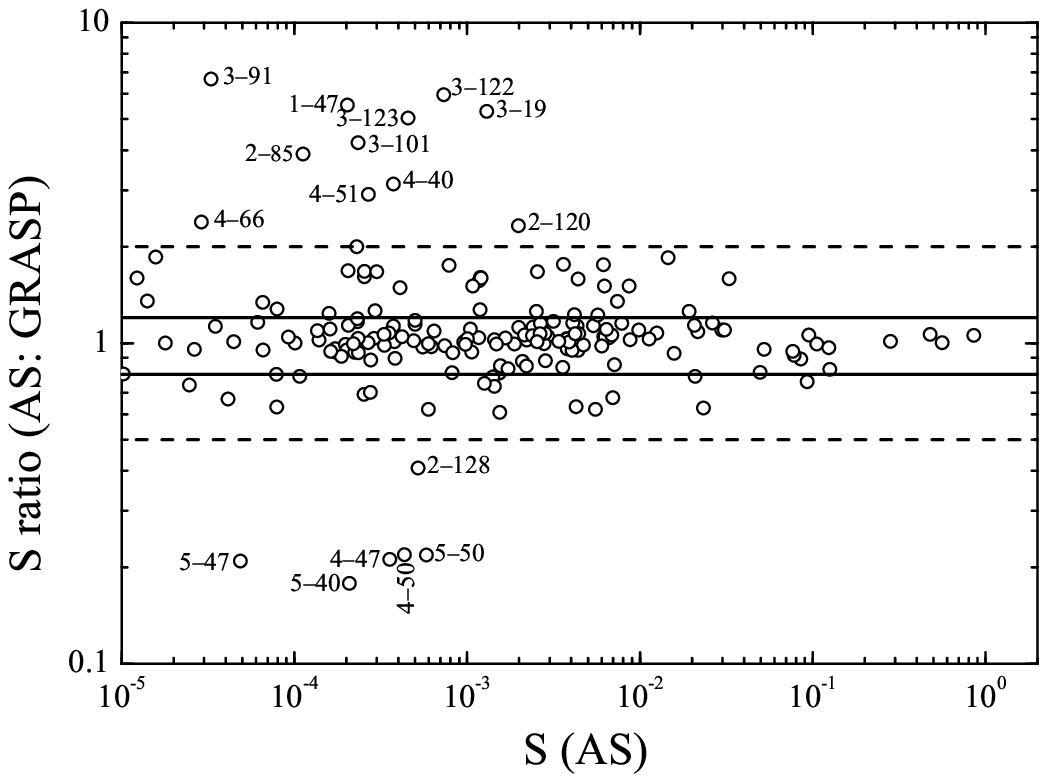}
\caption{Comparison of radiative decay rates, $A_{i,j}$, (all multipoles)
and electric dipole  line strengths ($S$) from the present {\sc autostructure} (AS)
calculation and the {\sc grasp} calculation by Aggarwal \& Keenan
(2007) for transitions to the lowest 5 levels. Solid and dashed
lines correspond to agreement within 20\% and a factor of 2,
respectively. Transitions with differences of more than a factor of 2
are marked by the transition label. (In the case of the rates,
labelling is restricted to strong rates: $\gtrsim 10^9~\rms^{-1}$.)}\label{fig_tran}
\end{figure*}

In the scatter plot of figure~\ref{fig_tran}, we compare radiative
decay rates ($A_{i,j}$ for the $i \leftarrow j$ transition)
for all electric and magnetic multipoles from our {\sc autostructure}
calculation with the {\sc grasp} data from AK07 for decays to the lowest
5-lying levels. Electric dipole line strengths are also shown.
Rates for some electric dipole transitions to the ground state are
listed in table~\ref{a-coeff}. For two-thirds of the transitions, 
the results of the two calculations agree to within 20\%, and 95\% of
transitions agree to within a factor of 2. We also note that there
are differences of more than a factor of 2, and up to an order
of magnitude for few transitions, which may be due to the
mis-match of the mapping of energy levels according to the parity,
total angular momentum and energy order scheme in the two
different calculations. Overall, the agreement is satisfactory.

\section{Scattering}
 For the present case of Fe$^{15+}$,
the resonance state configurations are of the form ${\rm
[2s,2p]}^{q-1}{\rm [3s,3p,3d]^2}nl$ (here $q=8$, $n\ge 3$), and
they can decay via the following channels
\begin{eqnarray}
{\rm [2s,2p]}^{q-1}{\rm [3s,3p,3d]^2}nl& \to & {\rm
[2s,2p]}^{q}{\rm [3s,3p,3d] + e}^-
\\
 & \to & {\rm [2s,2p]}^{q}nl+ {\rm e}^- \\
 & \to & {\rm [2s,2p]}^{q}{\rm [3s,3p,3d]^2} + h\nu \\
 & \to & {\rm [2s,2p]}^{q}{\rm [3s,3p,3d]}nl + h\nu\,.
\end{eqnarray}
The participator LM$n$ Auger pathway (1) scales as $n^{-3}$
and is automatically described in the $R$-matrix method by
the contribution to the close-coupling expansion from the right
hand side of (1).
However, the spectator LMM Auger pathway (2) is
independent of $n$ and only low-$n$ resonances ($n\leq3$ here) can be
included in the close-coupling expansion. But, the spectator Auger pathway
dominates for $n>3$. The last two channels, (3) and (4),
represent radiation damping. These Auger and radiation damping processes
reduce the resonant-enhancement of the excitation collision strengths and
can be expected to be especially important for inner-shell
transitions due to the large energy jump. 

For the Auger process, the
participator Auger channel can be included explicitly within the
$R$-matrix close-coupling expansion, whereas the spectator Auger
decay cannot easily be included for the higher-$n$ resonances as it
requires the inclusion of target states with $nl$ (with $n>3$) orbitals.
(So, only the Auger damping from $n=3$ resonances has been
included in the work of AK08). 

We employ the $R$-matrix intermediate-coupling frame
transformation (ICFT) method of Griffin \etal (1998) allowing for
both Auger-plus-radiation damping via the complex optical
potential, as described above. We used 25 continuum basis per
orbital angular momentum. Contributions from partial waves up to
$J$=12 were included in the exchange calculation. The
contributions from higher partial waves up to $J$=42 were included
via a non-exchange calculation. A `top-up' was used to complete
the partial collision strength sum over higher $J$-values by using
the Burgess sum rule (Burgess 1974) for dipole transitions and a
geometric series for the non-dipole transitions (Badnell \&
Griffin 2001). In the outer-region calculation, we used
an energy mesh step of 2$\times10^{-6}z^2$~Ryd through the resonance region
(from threshold to 72~Ryd), where $z$ is
the residual charge of the ion (15 in the present case). Beyond the
resonance region (from 72 to 450~Ryd), for the exchange
calculation, an energy step of 2$\times10^{-4}z^2~$Ryd was used.
For the non-exchange calculation, we used a step of
1$\times10^{-3}z^2~$Ryd over the entire energy range. The
calculation was carried-out up to an incident energy of 450~Ryd.
We used the infinite energy Born limits (non-dipole allowed) and
line-strengths (dipole-allowed) from {\sc autostructure} so that the
collision strengths could be interpolated at any desired energy
when Maxwell-averaging (see Whiteford \etal 2001).

Observed energies were used for the lowest 25-lying levels. For
those levels missing from the NIST database, we first derived the
mean value of differences between our level energies and the
corresponding NIST values for available levels of the ${\rm
2s^22p^53s3p}$ configuration, then we adjusted our calculated
level energies by this mean value. These observed and adjusted
energies are employed in the MQDT formula which converts from the
slowly-varying-with-energy unphysical $K$-matrix to the strongly
(resonant) energy-dependent physical $K$-matrix. This ensures that
Rydberg series of resonances converge on the observed thresholds.
In addition, low-lying (non-correlation) resonances can be
expected to be positioned accurately just above excitation
thresholds. A similar procedure has been
demonstrated to be very accurate in the study of dielectronic
recombination, where there is much precise experimental cross
section data with which to compare with~(see Savin \etal 2002, for example).

\begin{figure*}[t]
\includegraphics[angle=0,width=12.cm]{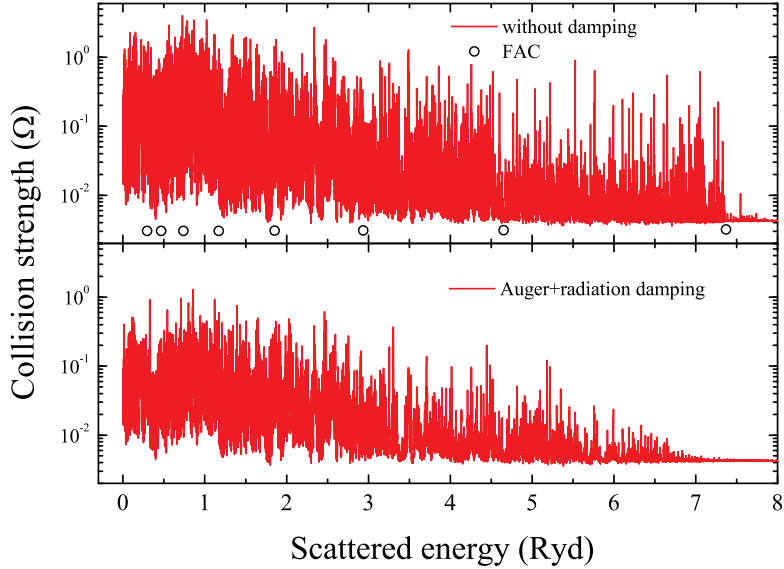}
\caption{ICFT $R$-matrix excitation collision strengths for the ${\rm
2s^22p^63s~^2S_{1/2}}$--${\rm 2s^22p^53s^2~^2P_{3/2}}$ transition,
without damping (top) and Auger-plus-radiation damping (bottom).
The circle symbols represent the results of our {\sc fac} DW
calculation. [{\it Colour online}]} \label{fig_cs}
\end{figure*}

In figure~\ref{fig_cs}, we show the collision strength of the
${\rm 2s^22p^63s~^2S_{1/2}}$--${\rm 2s^22p^53s^2~^2P_{3/2}}$
transition line, both without damping\footnote{Of course, $n=3$
Auger damping is still present here as it is intrinsic to the
close-coupling expansion.} as well as with Auger-plus-radiation
damping. The reduction due to the effect of Auger-plus-radiation
damping is very apparent on resonances, especially at higher
energies, and can be up to two orders of magnitude. Some
resonances are completely damped. The damping is dominated by far
($\sim 90$\%) by the Auger process for $n > 3$. We also performed
a distorted-wave (DW) calculation by using the {\sc fac} code with
the same configuration interactions as in our 134-level ICFT
$R$-matrix calculation. For this 1--6 transition, the DW data is
lower than the background value obtained from $R$-matrix by 25\%
at 8~Ryd.

Generally speaking, Maxwell-averaged effective collision strengths
($\Upsilon$) have a more extensive application than the ordinary
collision strengths ($\Omega$), in addition to the advantage of
a much smaller storage size. Test calculations show that the
effective collision strengths have converged (to within 1\% for
87\% of transitions) down to a temperature of 5.12$\times10^4$~K
on using an energy mesh step of 2$\times10^{-6}z^2~$Ryd.  At high
temperatures, effective collision strengths have converged on
using a much coarser mesh step of 5$\times10^{-6}z^2~$Ryd. So, in our
following work, we adopt an energy step of 2$\times10^{-6}z^2~$Ryd,
which is smaller than that adopted by AK08, by a factor of 2.

\section{Results and Discussions}
\subsection{Comparison of undamped results: ICFT $R$-matrix vs {\sc darc}}
\label{undamped} 
\begin{figure*}[t]
\hspace{-1.2cm}\includegraphics[angle=0,width=9.5cm]{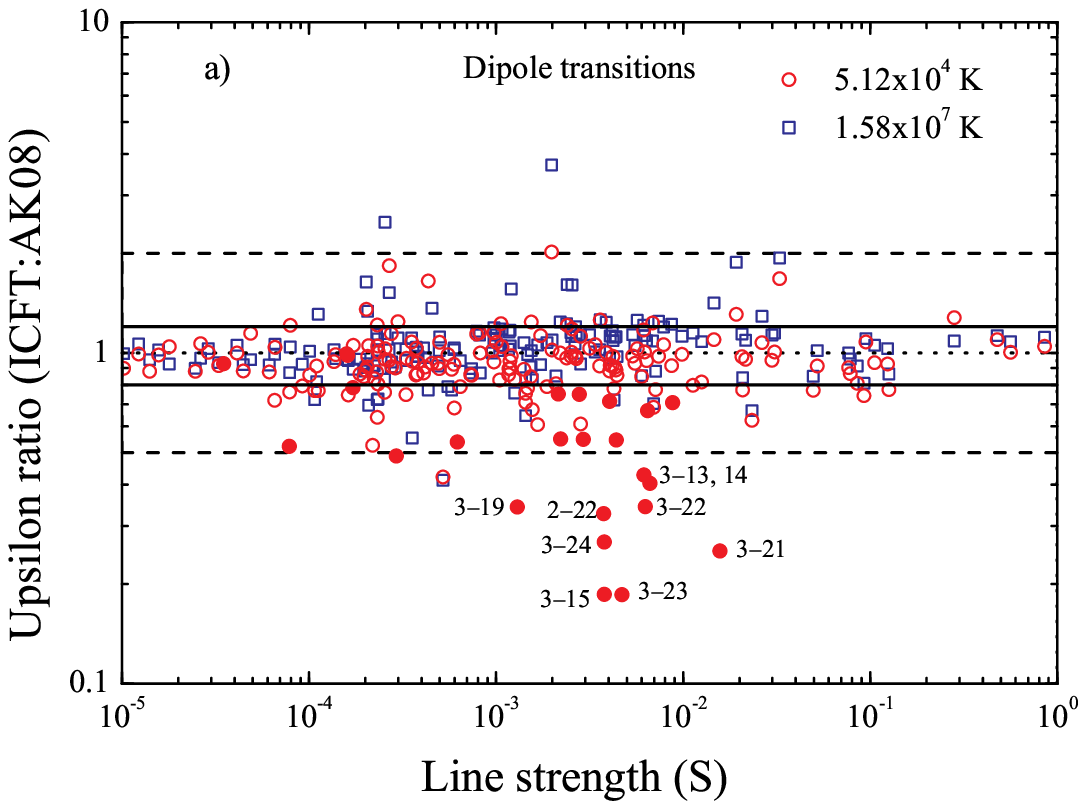}
\hspace{-1.2cm}\includegraphics[angle=0,width=9.5cm]{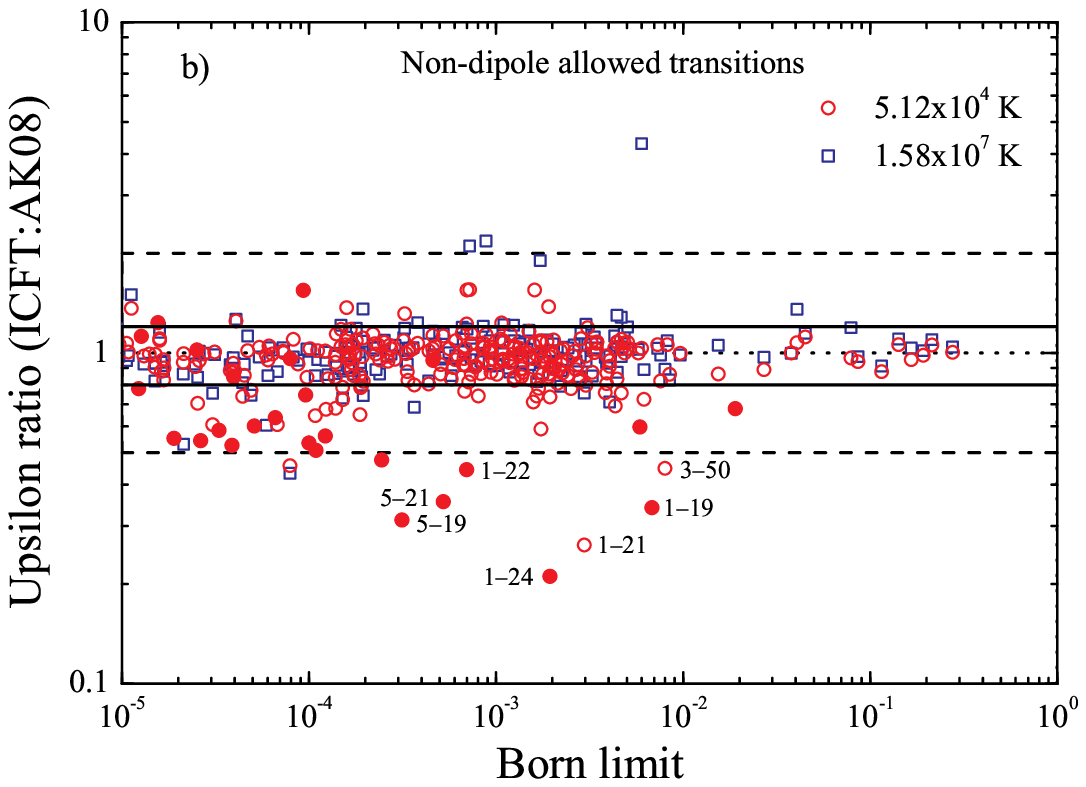}
\caption{Scatter plot of ratios of `undamped' effective collision
strengths from the present ICFT $R$-matrix calculation and the {\sc darc}
calculation of Aggarwal \& Keenan (2008) as a function of present AS a) line
strength ($S$) (for dipole transitions) and b) Born-limit (non-dipole allowed
transitions). ``$\circ$" and ``$\bullet$" symbols denote
transitions at 5.12$\times10^4$~K with threshold energy
differences between ICFT and AK08 calculations being less than and
greater than 0.2~Ryd, respectively. ``$\Box$" symbols:
corresponding results at 5.12$\times10^7$~K. Solid and dashed
lines correspond to agreement within 20\% and a factor of 2,
respectively. Dotted lines mark where the ratios agree. [{\it
Colour online}]} \label{fig_undp-darc}
\end{figure*}

We make comparison with the results of AK08
calculated by using {\sc darc}. We make contrasting comparisons at a low
temperature (5.12$\times10^4$~K) and a high one
(1.58$\times10^7$~K) as shown in figure~\ref{fig_undp-darc}, in
which transitions from the lowest 5-lying levels to all higher
levels (total 655 transitions) are plotted for dipole allowed (211) and
non-dipole allowed (426) transitions. In intermediate coupling,
spin--orbit mixing means that very few transitions that were
forbidden in $LS$-coupling remain so. Instead, they have small but
non-zero line strengths or infinite energy Born limits. Indeed,
only 18 transitions from the lowest 5-lying levels are strictly
forbidden according to this definition (and are not shown in
figure~4).

For most transitions, our undamped ICFT $R$-matrix results agree
with the {\sc darc} ones of AK08, to within 20\% over the entire
temperature range. At the low temperature (5.12$\times10^4$~K),
there are 35.5\% of dipole and 20.7\% of non-dipole allowed
transitions with a difference of over 20\%. This difference
decreases to 25.1\% of dipole and 13.8\% of non-dipole allowed
transitions at the high temperature (1.58$\times10^7$~K) at the
low temperature (5.12$\times10^4$~K). Here, for dipole
transitions, we find that there is a strong correlation between
the ratio of the ICFT to {\sc darc} $\Upsilon$ values and the
ratio of the AS to GRASP line strengths. The ICFT/{\sc darc}
agreement for non-dipole allowed transitions should also be
strongly correlated with the atomic structure
--- this time for the infinite energy Born limit, but we do not have
such results for {\sc grasp}.

In figure~4, we identify a group of dipole and non-dipole allowed
transitions (see table~\ref{diff_tran}) for which the ratio of
line strengths (electric dipole only) is close to unity but the {\sc darc}
effective collision strengths are systematically larger than those
obtained from ICFT, at the lower temperature. This is probably due
to the smaller excitation energies used by AK08 (recall, we
adjusted to observed) which means that there are additional
resonances present at lower energies/temperatures in the AK08
data. This can be tested indirectly by looking at excitations to
higher levels, which also have strong resonant contributions. For
example, for the 1--28 transition (not shown) with a threshold
energy difference of 0.041Ryd, the $\Upsilon$ values are
7.58$\times10^{-3}$ and 7.61$\times10^{-3}$, respectively, at the
low temperature.

\begin{figure*}[t]
\hspace{-1.cm}\includegraphics[angle=0,width=9.5cm]{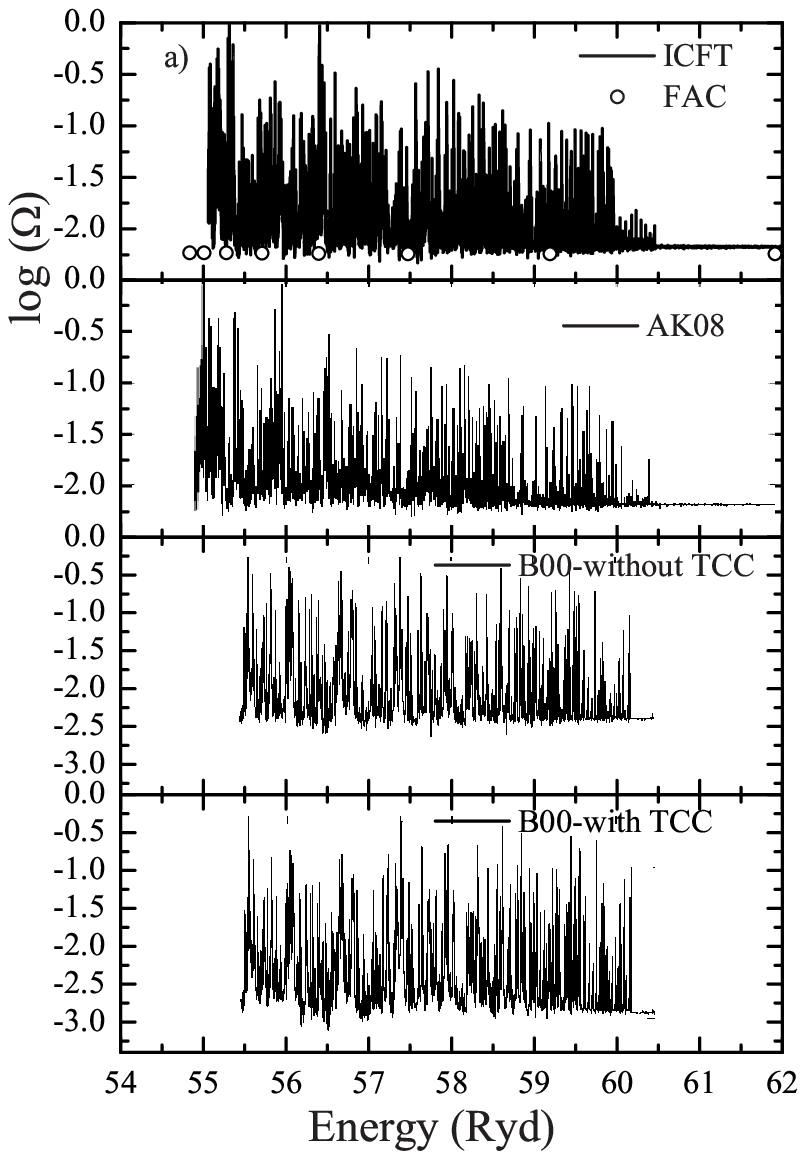}
\hspace{-1.cm}\includegraphics[angle=0,width=9.5cm]{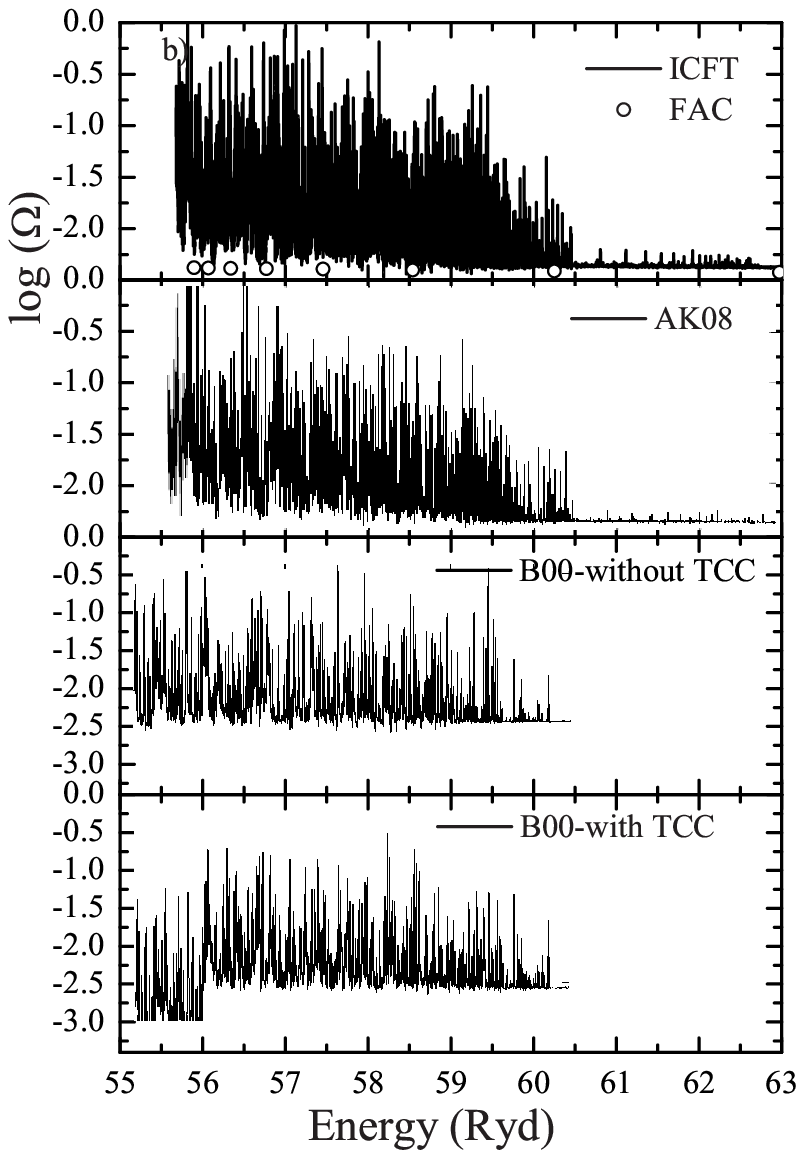}
\caption{Collision strengths ($\Omega$) from the  ICFT $R$-matrix (present
work), {\sc darc} (AK08) and {\sc jajom} (Bautista (2000): B00)
calculations. a): for 1--15 (${\rm 2s^22p^63s~^2S_{1/2}}$--${\rm
2s^22p^53s3p~^2S_{1/2}}$) transition line. b): 1--21 (${\rm
2s^22p^63s~^2S_{1/2}}$--${\rm 2s^22p^53s3p~^2D_{5/2}}$) transition
line. ``$\circ$" denote DW values obtained from {\sc fac} (present
work). } \label{fig_cs_undp_darc}
\end{figure*}

AK08 selected the 1--15 and 1--21 transitions to reveal
inadequacies of term-coupling via the {\sc jajom} code, as used by
Bautista (2000), which results in the sudden increase and/or
decrease of background collision strengths when relativistic
effects are included for some transitions (see figure~3 in
Bautista (2000) and the bottom two panels in
figure~\ref{fig_cs_undp_darc}). This is exactly the same
inadequacy demonstrated originally by Griffin \etal (1998) when
they introduced the ICFT $R$-matrix method to solve the problem, without
resorting to a full Breit--Pauli (or Dirac) calculation. AK08
conclude that this inadequacy is the reason for the large
discrepancies between the results of their two calculations. In order
to illustrate the inadequacy of the {\sc jajom} method and the
overestimation of AK08 at the lower temperature for some
transitions (see filled circles in figure~\ref{fig_undp-darc}-a),
we compare the underlying collision strengths for 1--15 and 1--21
transitions in figure~\ref{fig_cs_undp_darc}. We see that the
background does not shift down and no sudden jumps appear in the
present ICFT results, in contrast to that seen from {\sc jajom} --- see
the bottom two panels of figure~\ref{fig_cs_undp_darc}. The
background of ICFT $\Omega$-values show excellent agreement with
the {\sc darc} ones. Our DW results obtained from {\sc fac} are also overlapped,
showing an excellent agreement with the background result of the two
$R$-matrix calculations. Additionally, the resonance structures in
the two $R$-matrix calculations basically agree with each other.
However, because the energy of ${\rm 2s^22p^53s3p~^2S_{1/2}}$
(15-) and ${\rm ^2D_{5/2}}$ (21-) levels of the AK08's data are
lower than the observed values which we use, by $\approx 0.2$~Ryd,
resonances around this region appear in the work of AK08, as shown
in figure~\ref{fig_cs_undp_darc}. So, their results are probably
somewhat of an overestimate of the effective collision strengths
at lower temperatures.

\begin{figure*}[t]
\includegraphics[angle=0,width=10.5cm]{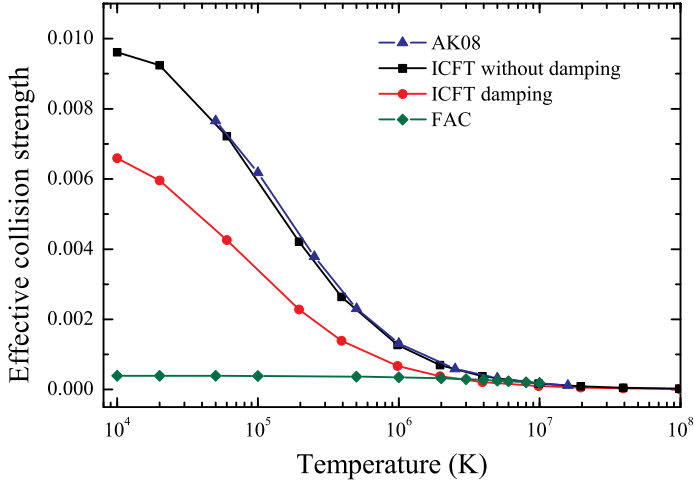}
\caption{Comparison of the effective collision strengths for the
${\rm 2s^22p^63s~^2S_{1/2}}$ --- ${\rm 2s^22p^53p^2~^4P_{5/2}}$
transition $(1-28)$. ICFT $R$-matrix and {\sc fac} DW are present results.
AK08 denotes the {\sc darc} results of Aggarwal
\& Keenan (2008). [{\it Colour online}]} \label{fig_eff}
\end{figure*}

\begin{figure*}[h]
\includegraphics[angle=0,width=9.5cm]{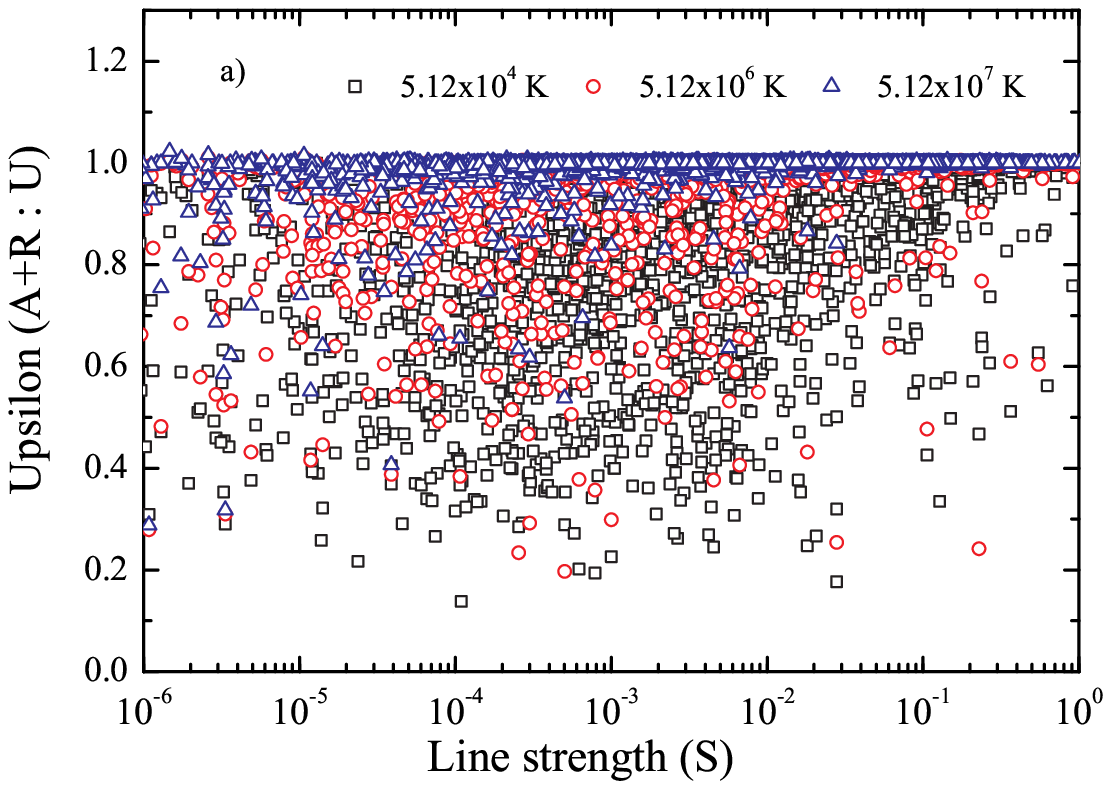}
\hspace{-1.0cm}\includegraphics[angle=0,width=9.5cm]{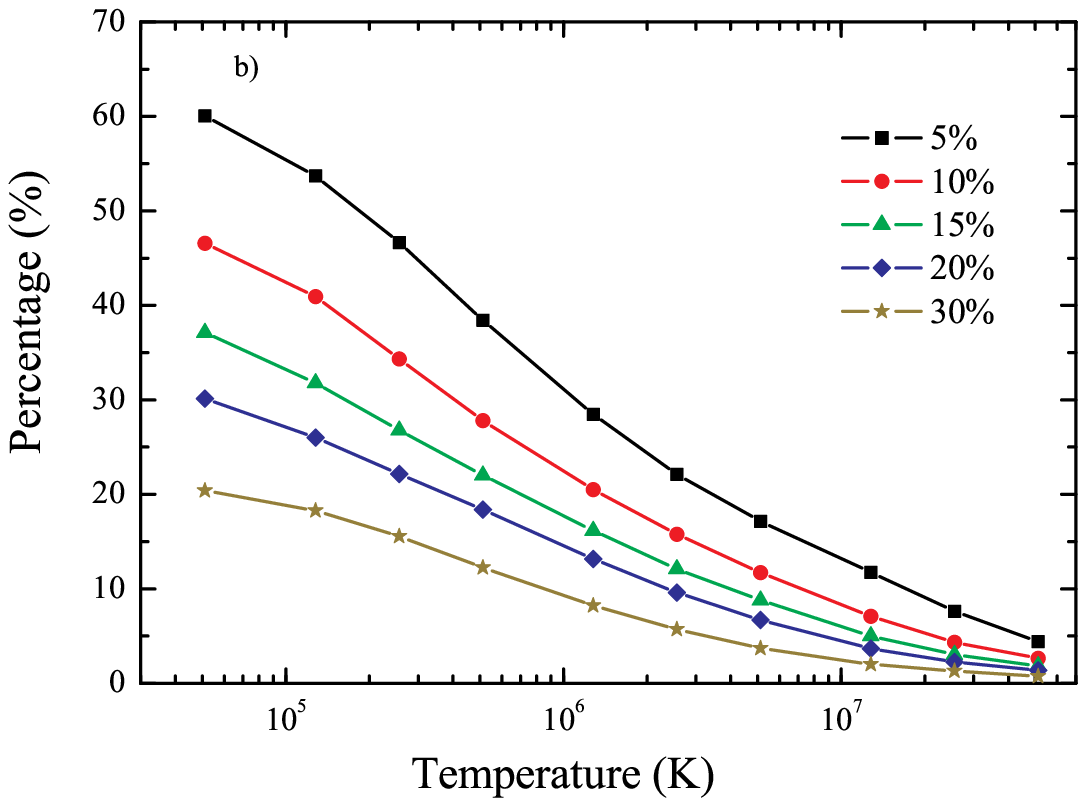} \vspace{-1.0cm}\\
\hspace{-1.5cm}\includegraphics[angle=0,width=9.5cm]{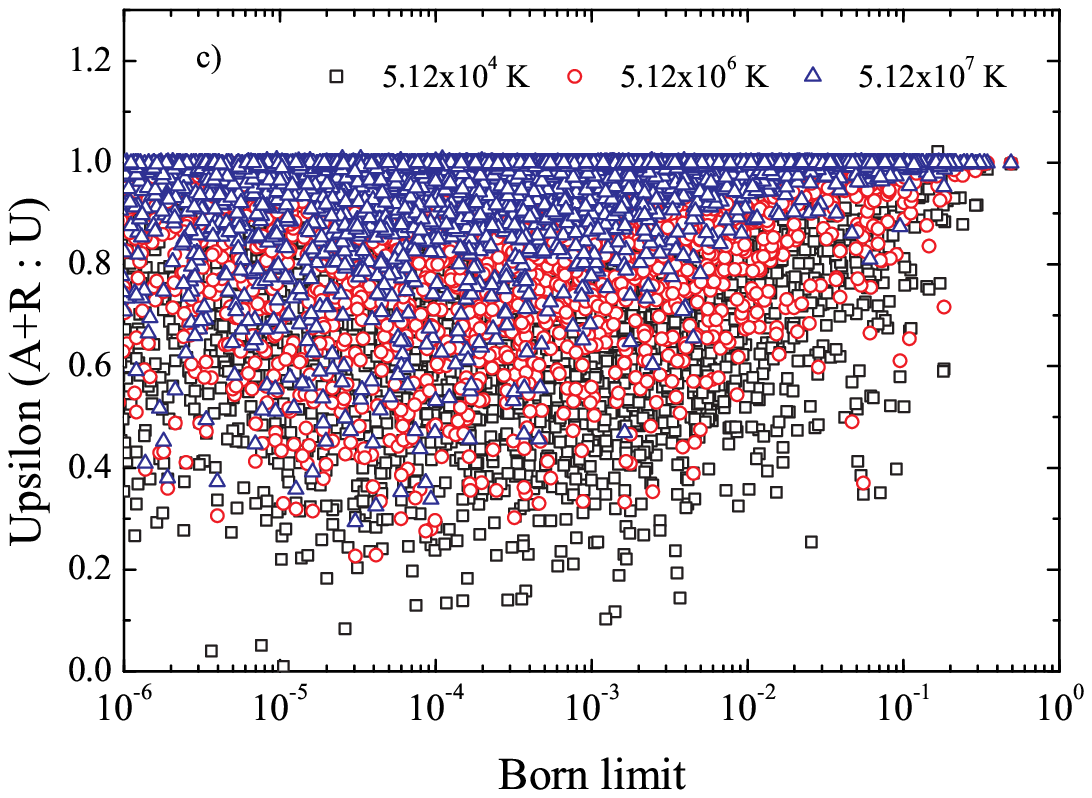}
\hspace{-1.0cm}\includegraphics[angle=0,width=9.5cm]{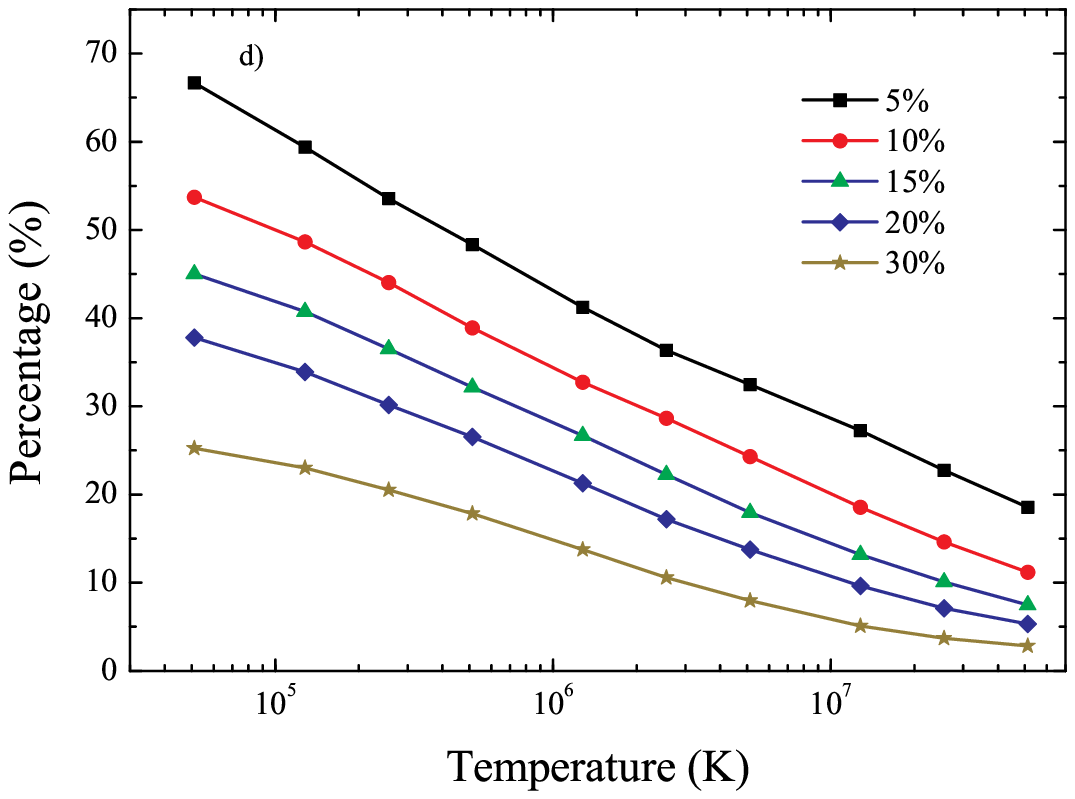} \vspace{-1.0cm}\\
\hspace{-1.5cm}\includegraphics[angle=0,width=9.5cm]{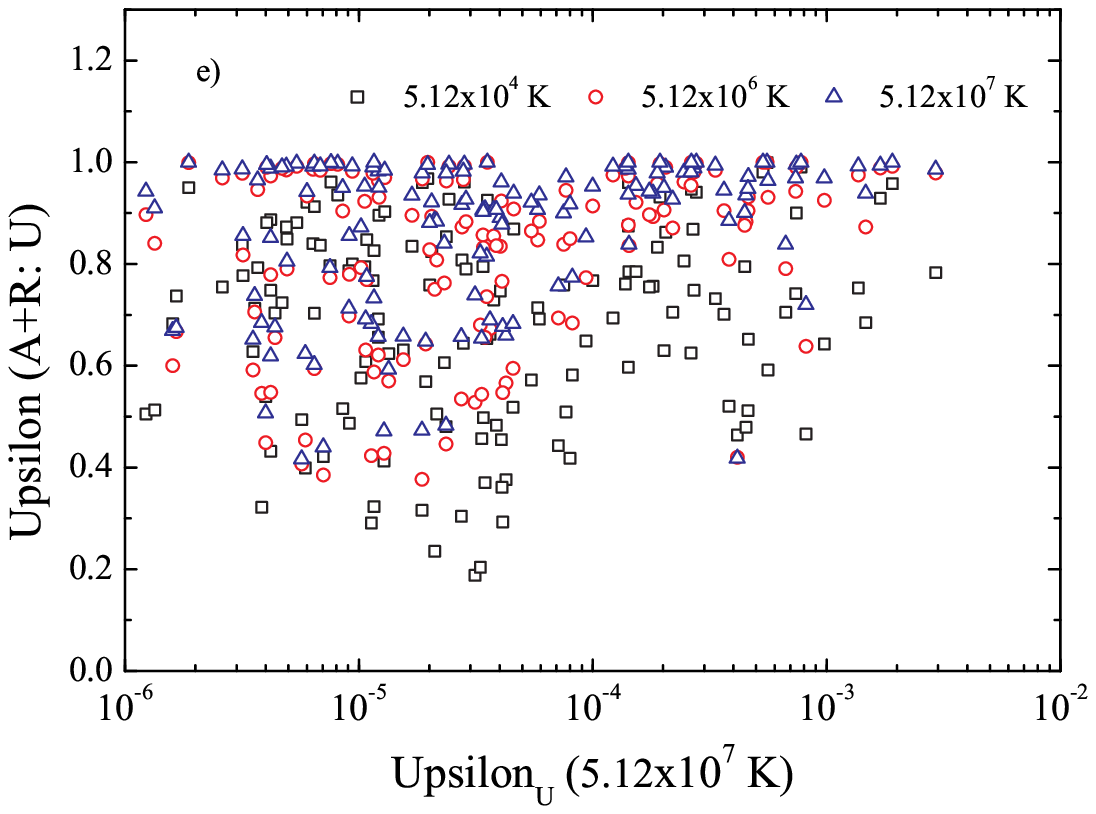}
\hspace{-1.0cm}\includegraphics[angle=0,width=9.5cm]{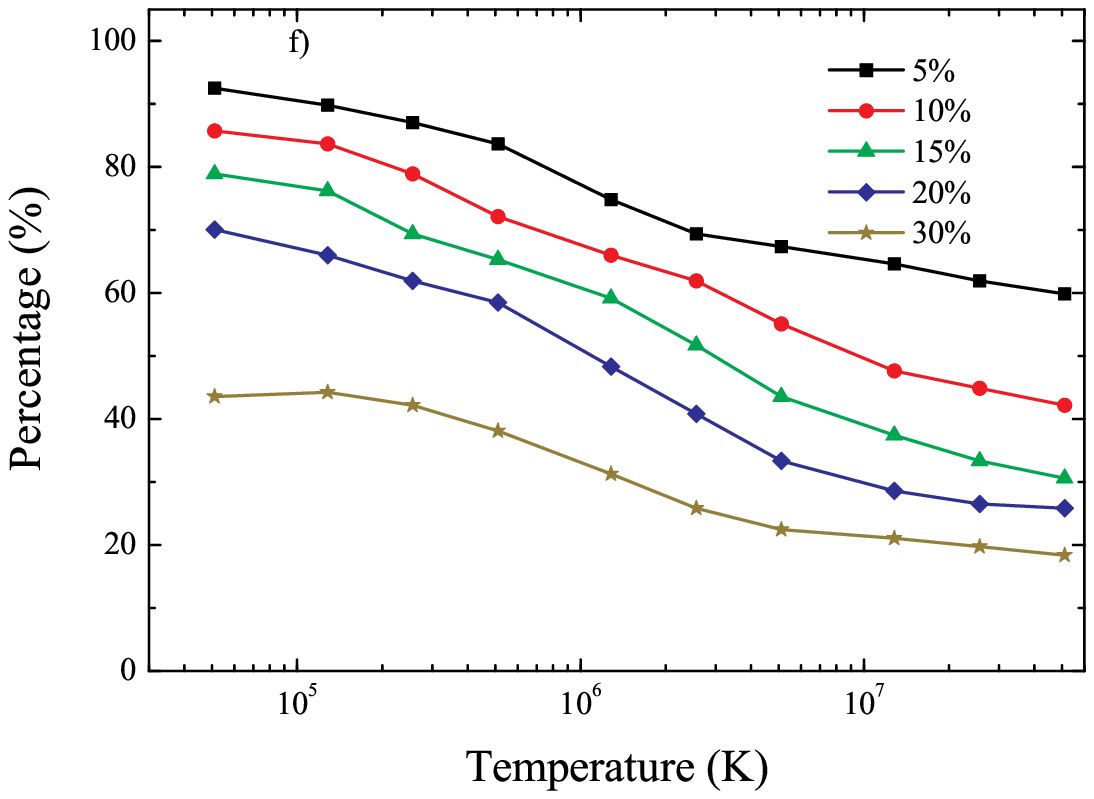}
\caption{Left-hand panels: Scatter plots showing the ratio of the
effective collision strengths ($\Upsilon$) with
Auger-plus-radiation damping (Upsilon(A+R)) to without damping
(Upsilon$_{\rm U}$) as a function of a) line strength, c) infinite temperature
Born limit, e) undamped $\Upsilon$ at the highest tabulated temperature
(5.12$\times10^7$~K), for dipole, non-dipole allowed and forbidden
transitions, respectively. Right-hand panels: percentage of
corresponding transitions where the effect of damping exceeds 5\%,
10\%, 15\%, 20\% and 30\%. [{\it Colour online}]}
\label{fig_dpeff}
\end{figure*}

\subsection{Comparison of ICFT $R$-matrix results: damped vs undamped}
\label{damped}

Figure~\ref{fig_eff} shows the results of several calculations of the
effective collision strength $(\Upsilon)$ for the
${\rm 2s^22p^63s~^2S_{1/2}}$ --- ${\rm 2s^22p^53p^2~^4P_{5/2}}$
transition $(1-28)$.
They demonstrate the physics we seek to describe: firstly, on comparing
$R$-matrix results with our DW ones obtained with {\sc fac}, we see that
the resonant enhancement is about a factor of 8 at $2\times 10^4$~K
(typical of where Fe$^{15+}$ is abundant in photoionized plasmas);
secondly, there is close agreement between our present undamped ICFT $R$-matrix
results and the {\sc darc} $R$-matrix ones of AK08; finally, Auger-plus-radiation
damping lowers the resonance enhanced results by nearly a factor of 2, again
at $2\times 10^4$~K.

The widespread effect of Auger-plus-radiation damping is
illustrated via a scatter plot of the ratios of
damped to undamped $\Upsilon$ values  for dipole
(figure~\ref{fig_dpeff}-a), non-dipole allowed
(figure~\ref{fig_dpeff}-c) and forbidden
(figure~\ref{fig_dpeff}-e) transitions. We see that the reduction
at the low temperature (5.12$\times10^4$~K) can be up to a factor
of 3 for a few (1.3\%) dipole transitions. The effect reduces
with increasing of temperature and is less than 10\% for 97.5\% of
these transitions at the high temperature
(5.12$\times10^7$~K). An illustrative way to quantify the information in the
scatter plot is to count how many transitions differ by more than
a given amount. In figure~\ref{fig_dpeff}-b, we show the
percentage of each class of transition where the damping
effect is at least 5\%, 10\%, 15\%, 20\% and 30\%. About 20\% of
dipole transitions show a damping effect of more than 30\% at
$5.12\times10^4$~K. At higher temperatures, the damping becomes a
smaller effect --- less than 6\% of dipole transitions show a
$>$30\% effect at 2.56$\times10^6$~K, for example. For
non-dipole allowed transitions (see figure~\ref{fig_dpeff}-c), the
damping effect can be up to a factor 3 for some transitions (2.1\%)
with a Born limit between 10$^{-4}$ and 10$^{-2}$
at the low temperature (5.12$\times10^4$~K). The effect reduces to
less than 10\% for 88.7\% of these transitions at the high
temperature (5.12$\times10^7$~K). Counting statistics (see
figure~\ref{fig_dpeff}-d) reveals that $\approx$25\% of non-dipole allowed
transitions show a reduction more than 30\% at the low temperature.
There are only a few forbidden transitions (1.6\% of the 8911
transitions in total). The damping effect is stronger for weaker
excitations, see figure~\ref{fig_dpeff}-e. About 44\% of forbidden
transitions show a damping effect over 30\% at the low temperature
(5.12$\times10^4$~K) --- see figure~\ref{fig_dpeff}-f. At the
high temperature, the percentage is still 40\% of
forbidden transitions with damping over 10\%. This value is
significantly higher than that for dipole (2.5\%) and non-dipole allowed
(11.3\%) transitions. We also note that the forbidden transitions
are affected over a wider range of electron temperatures.

Finally, in table~\ref{upsdamp}, the undamped and damped effective
collision strengths  are given  for excitations from the ground
level at three temperatures of 5.12$\times10^4$, 5.12$\times10^5$,
and 5.12$\times10^6$~K. The full set of data (energy levels,
radiative rates and effective collision strengths) are make
available through different archives and databases (the Oak Ridge
Controlled Fusion Atomic Data Center (CFADC)\footnote
{http://www-cfadc.phy.ornl.gov/data\_and\_codes} in the ADAS {\it
adf04} format (Summers 2004),
ADAS\footnote{http://www.adas.ac.uk/} and
CHIANTI\footnote{http://www.chianti.rl.ac.uk/}).

\section{Conclusions}
The level-resolved inner-shell electron-impact excitation of
Fe$^{15+}$ has been studied via the intermediate coupling frame
transformation {\it R}-matrix method which can allow for the
inclusion of Auger-plus-radiation damping of such
resonantly-excited states. The 134 levels belonging to the
configurations ${\rm 2s^22p^63}l$, ${\rm 2s^22p^53s3}l$ ($l$=s,
p and d), ${\rm 2s^22p^53p^2}$ and ${\rm 2s^22p^53p3d}$ were
included in both the target configuration and close-coupling
expansions. A comparison of energy levels and radiative rates
with those of Aggarwal \& Keenan (2007) reveals the target
structures to be comparable, and so form the basis for comparison
with the excitation data of Aggarwal \& Keenan (2008).

The results of our undamped ICFT $R$-matrix calculation agree well with the
undamped {\sc darc} effective collision strengths of Aggarwal \& Keenan
(2008) for most excitations. For a few transitions, their results
are higher than ours by a factor of two at low temperatures. This
is probably due to their use of smaller theoretical transition energies
than ours, which were adjusted to the observed values. When
Auger-plus-radiation damping is included, our results are
systematically smaller than those of Aggarwal \& Keenan (2008).
Moreover, the reduction can be up to a factor of 3 for some
transitions. The number of transitions where the reduction of
$\Upsilon$ exceeds 20\% occupies 30.2\%, 37.7\% and 70.7\% of
dipole, non-dipole allowed and forbidden transitions, respectively,
at the low temperatures typical of where Fe$^{15+}$ is abundant in
photoionized plasmas.

In summary, Auger-plus-radiation damping plays an
important role on the electron-impact excitation of inner-shell
transitions. Thus, for many transitions, the results of previous undamped inner-shell
calculations overestimate the effective collision strengths significantly.

\section{Acknowledgments}
The work of the UK APAP program is funded by the UK STFC under
grant no. PP/E001254/1 with the University of Strathclyde. One of
us (GY) would like to thank M. C. Witthoeft for some helpful
discussions.


\clearpage
\pagestyle{empty}
\Table{Energy levels (Ryd), and differences, for
the ${\rm 2s^22p^63}l$, ${\rm 2s^22p^53s3}l$ ($l$=s,~p~and~d),
${\rm 2s^22p^53p^2}$ and ${\rm 2s^22p^53p3d}$ configurations of
Fe$^{15+}$. \label{energy}}
\br
Index & Configuration & $^{2S+1}L_J$ & NIST$^\rma$ &
AS$^\rmb$ & GRASP$^\rmc$ & FAC$^\rmd$ & AS--NIST & GRASP\\
& & & & & & & &--NIST \\
\mr
\01  & $    2\rms^2 2\rmp^6 3\rms  $ & $^2\rmS_{1/2} $ & \00.00000  &\00.00000    &\00.00000  &\00.00000  & 0.00000   &  0.00000  \\
\02  & $    2\rms^2 2\rmp^6 3\rmp  $ & $^2\rmP_{1/2} $ & \02.52598  &\02.55904    &\02.56618  &\02.54749  & 0.03306   &  0.04020  \\
\03  & $    2\rms^2 2\rmp^6 3\rmp  $ & $^2\rmP_{3/2} $ & \02.71688  &\02.76660    &\02.75486  &\02.73695  & 0.04972   &  0.03798  \\
\04  & $    2\rms^2 2\rmp^6 3\rmd  $ & $^2\rmD_{3/2} $ & \06.15562  &\06.17619    &\06.19591  &\06.16930  & 0.02057   &  0.04029  \\
\05  & $    2\rms^2 2\rmp^6 3\rmd  $ & $^2\rmD_{5/2} $ & \06.18209  &\06.21736    &\06.22081  &\06.19488  & 0.03527   &  0.03872  \\
\06  & $   2\rms^2 2\rmp^5 3\rms^2 $ & $^2\rmP_{3/2} $ & 52.60745  & 52.34842   & 52.31794 & 52.41368 &  \-0.25903  &  \-0.28951 \\
\07  & $   2\rms^2 2\rmp^5 3\rms^2 $ & $^2\rmP_{1/2} $ & 53.51871  & 53.28728   & 53.24028 & 53.33365 &  \-0.23143  &  \-0.27843 \\
\08  & $ 2\rms^2 2\rmp^5 3\rms 3\rmp  $ & $^4\rmS_{3/2} $ &           & 54.11760   & 54.08374 & 54.17677 &            &           \\
\09  & $ 2\rms^2 2\rmp^5 3\rms 3\rmp  $ & $^4\rmD_{5/2} $ & 54.51199  & 54.36541   & 54.32940 & 54.42627 &  \-0.14658  &  \-0.18259 \\
 10  & $ 2\rms^2 2\rmp^5 3\rms 3\rmp  $ & $^4\rmD_{7/2} $ &           & 54.41366   & 54.38145 & 54.47648 &            &           \\
 11  & $ 2\rms^2 2\rmp^5 3\rms 3\rmp  $ & $^2\rmP_{3/2} $ &           & 54.42411   & 54.38938 & 54.48968 &            &           \\
 12  & $ 2\rms^2 2\rmp^5 3\rms 3\rmp  $ & $^2\rmP_{1/2} $ & 54.68514  & 54.55154   & 54.52045 & 54.61949 &  \-0.13360  &  \-0.16469 \\
 13  & $ 2\rms^2 2\rmp^5 3\rms 3\rmp  $ & $^4\rmP_{5/2} $ &           & 54.65336   & 54.62489 & 54.72504 &            &           \\
 14  & $ 2\rms^2 2\rmp^5 3\rms 3\rmp  $ & $^2\rmD_{3/2} $ & 54.79442  & 54.65685   & 54.62869 & 54.73060 &  \-0.13757  &  \-0.16573 \\
 15  & $ 2\rms^2 2\rmp^5 3\rms 3\rmp  $ & $^2\rmS_{1/2} $ & 55.05876  & 54.87978   & 54.84442 & 54.93987 &  \-0.17898  &  \-0.21434 \\
 16  & $ 2\rms^2 2\rmp^5 3\rms 3\rmp  $ & $^4\rmD_{1/2} $ & 55.35947  & 55.26494   & 55.21306 & 55.30885 &  \-0.09453  &  \-0.14641 \\
 17  & $ 2\rms^2 2\rmp^5 3\rms 3\rmp  $ & $^4\rmD_{3/2} $ & 55.48705  & 55.35890   & 55.30739 & 55.40441 &  \-0.12815  &  \-0.17966 \\
 18  & $ 2\rms^2 2\rmp^5 3\rms 3\rmp  $ & $^4\rmP_{1/2} $ & 55.48705  & 55.37835   & 55.33684 & 55.42804 &  \-0.10870  &  \-0.15021 \\
 19  & $ 2\rms^2 2\rmp^5 3\rms 3\rmp  $ & $^4\rmP_{3/2} $ & 55.55084  & 55.47899   & 55.43583 & 55.52946 &  \-0.07185  &  \-0.11501 \\
 20  & $ 2\rms^2 2\rmp^5 3\rms 3\rmp  $ & $^2\rmD_{5/2} $ &           & 55.47904   & 55.44467 & 55.53084 &            &           \\
 21  & $ 2\rms^2 2\rmp^5 3\rms 3\rmp  $ & $^2\rmD_{5/2} $ & 55.67842  & 55.54816   & 55.50469 & 55.60098 &  \-0.13026  &  \-0.17373 \\
 22  & $ 2\rms^2 2\rmp^5 3\rms 3\rmp  $ & $^2\rmP_{3/2} $ & 55.85156  & 55.61425   & 55.58697 & 55.67445 &  \-0.23731  &  \-0.26459 \\
 23  & $ 2\rms^2 2\rmp^5 3\rms 3\rmp  $ & $^2\rmP_{1/2} $ &           & 56.29895   & 56.25643 & 56.34356 &            &           \\
 24  & $ 2\rms^2 2\rmp^5 3\rms 3\rmp  $ & $^2\rmD_{3/2} $ & 56.65347  & 56.45616   & 56.40948 & 56.49089 &  \-0.19731  &  \-0.24399 \\
 25  & $ 2\rms^2 2\rmp^5 3\rms 3\rmp  $ & $^2\rmS_{1/2} $ & 57.10911  & 57.01687   & 56.99370 & 57.08592 &  \-0.09224  &  \-0.11541 \\
 26  & $   2\rms^2 2\rmp^5 3\rmp^2 $ & $^4\rmP_{3/2} $ &           & 57.10417   & 57.06704 & 57.15630 &            &           \\
 27  & $   2\rms^2 2\rmp^5 3\rmp^2 $ & $^2\rmP_{1/2} $ &           & 57.10468   & 57.06488 & 57.15917 &            &           \\
 28  & $   2\rms^2 2\rmp^5 3\rmp^2 $ & $^4\rmP_{5/2} $ &           & 57.17352   & 57.13944 & 57.22799 &            &           \\
 29  & $   2\rms^2 2\rmp^5 3\rmp^2 $ & $^2\rmF_{7/2} $ &           & 57.19233   & 57.14806 & 57.24427 &            &           \\
 30  & $   2\rms^2 2\rmp^5 3\rmp^2 $ & $^2\rmP_{3/2} $ &           & 57.26544   & 57.22768 & 57.32010 &            &           \\
 31  & $   2\rms^2 2\rmp^5 3\rmp^2 $ & $^2\rmD_{5/2} $ &           & 57.38943   & 57.35118 & 57.44727 &            &           \\
 32  & $   2\rms^2 2\rmp^5 3\rmp^2 $ & $^2\rmD_{3/2} $ &           & 57.42587   & 57.39127 & 57.48631 &            &           \\
 33  & $   2\rms^2 2\rmp^5 3\rmp^2 $ & $^4\rmP_{1/2} $ &           & 57.44009   & 57.40547 & 57.49546 &            &           \\
 34  & $   2\rms^2 2\rmp^5 3\rmp^2 $ & $^4\rmD_{7/2} $ &           & 57.45369   & 57.42105 & 57.51216 &            &           \\
 35  & $   2\rms^2 2\rmp^5 3\rmp^2 $ & $^4\rmD_{5/2} $ &           & 57.46957   & 57.43729 & 57.53005 &            &           \\
 36  & $   2\rms^2 2\rmp^5 3\rmp^2 $ & $^4\rmD_{1/2} $ &           & 57.93100   & 57.87458 & 57.96558 &            &           \\
 37  & $   2\rms^2 2\rmp^5 3\rmp^2 $ & $^4\rmS_{3/2} $ &           & 57.96240   & 57.91773 & 58.01194 &            &           \\
 38  & $ 2\rms^2 2\rmp^5 3\rms 3\rmd  $ & $^4\rmP_{1/2} $ &           & 58.03132   & 57.93888 & 58.02976 &            &           \\
 39  & $ 2\rms^2 2\rmp^5 3\rms 3\rmd  $ & $^4\rmP_{3/2} $ &           & 58.10048   & 58.00538 & 58.09616 &            &           \\
 40  & $   2\rms^2 2\rmp^5 3\rmp^2 $ & $^2\rmF_{5/2} $ &           & 58.18227   & 58.12751 & 58.18495 &            &           \\
 41  & $   2\rms^2 2\rmp^5 3\rmp^2 $ & $^4\rmD_{3/2} $ &           & 58.19672   & 58.09555 & 58.20798 &            &           \\
 42  & $ 2\rms^2 2\rmp^5 3\rms 3\rmd  $ & $^4\rmF_{9/2} $ &           & 58.19737   & 58.14156 & 58.21959 &            &           \\
 43  & $ 2\rms^2 2\rmp^5 3\rms 3\rmd  $ & $^4\rmP_{5/2} $ & 58.25730  & 58.21984   & 58.11794 & 58.23081 &  \-0.03746  &  \-0.13936 \\
\br
\end{tabular}
\end{indented}
\end{table}
\setcounter{table}{0}
\Table{-continued}
\br
Index & Configuration & $^{2S+1}L_J$ & NIST & AS & GRASP & FAC & AS--NIST & GRASP\\
& & & & & & & &--NIST \\
\mr
 44  & $ 2\rms^2 2\rmp^5 3\rms 3\rmd  $ & $^4\rmF_{7/2} $ &           & 58.24983   & 58.16102 & 58.24724 &            &          \\
 45  & $   2\rms^2 2\rmp^5 3\rmp^2 $ & $^2\rmS_{1/2} $ &           & 58.26394   & 58.21189 & 58.30063 &            &          \\
 46  & $ 2\rms^2 2\rmp^5 3\rms 3\rmd  $ & $^4\rmF_{5/2} $ & 58.37577  & 58.32413   & 58.24386 & 58.32935 &  \-0.05164  & \-0.13191 \\
 47  & $   2\rms^2 2\rmp^5 3\rmp^2 $ & $^2\rmD_{3/2} $ &           & 58.39817   & 58.36742 & 58.42791 &            &          \\
 48  & $ 2\rms^2 2\rmp^5 3\rms 3\rmd  $ & $^2\rmD_{3/2} $ &           & 58.42714   & 58.34352 & 58.45325 &            &          \\
 49  & $ 2\rms^2 2\rmp^5 3\rms 3\rmd  $ & $^4\rmD_{7/2} $ & 58.52157  & 58.47411   & 58.39564 & 58.47794 &  \-0.04746  & \-0.12593 \\
 50  & $   2\rms^2 2\rmp^5 3\rmp^2 $ & $^2\rmD_{5/2} $ &           & 58.48076   & 58.43827 & 58.50728 &            &          \\
 51  & $   2\rms^2 2\rmp^5 3\rmp^2 $ & $^2\rmP_{3/2} $ & 58.54891  & 58.50036   & 58.42242 & 58.53072 &  \-0.04855  & \-0.12649 \\
 52  & $ 2\rms^2 2\rmp^5 3\rms 3\rmd  $ & $^2\rmF_{5/2} $ &           & 58.50383   & 58.45493 & 58.53880 &            &          \\
 53  & $ 2\rms^2 2\rmp^5 3\rms 3\rmd  $ & $^2\rmP_{1/2} $ & 58.53068  & 58.54673   & 58.48029 & 58.56573 &  0.01605   & \-0.05039 \\
 54  & $ 2\rms^2 2\rmp^5 3\rms 3\rmd  $ & $^2\rmP_{3/2} $ & 58.64915  & 58.72397   & 58.66234 & 58.74315 &  0.07482   & 0.01319  \\
 55  & $ 2\rms^2 2\rmp^5 3\rms 3\rmd  $ & $^4\rmD_{1/2} $ &           & 58.81395   & 58.76090 & 58.83584 &            &          \\
 56  & $ 2\rms^2 2\rmp^5 3\rms 3\rmd  $ & $^4\rmD_{3/2} $ & 58.98632  & 59.11274   & 59.04324 & 59.12033 &  0.12642   & 0.05692  \\
 57  & $ 2\rms^2 2\rmp^5 3\rms 3\rmd  $ & $^4\rmF_{3/2} $ &           & 59.22437   & 59.14077 & 59.20637 &            &          \\
 58  & $ 2\rms^2 2\rmp^5 3\rms 3\rmd  $ & $^2\rmF_{7/2} $ & 58.73116  & 59.22499   & 59.13497 & 59.22747 &  0.49383   & 0.40381  \\
 59  & $ 2\rms^2 2\rmp^5 3\rms 3\rmd  $ & $^4\rmD_{5/2} $ & 59.25058  & 59.25061   & 59.15303 & 59.23382 &  0.00003   & \-0.09755 \\
 60  & $ 2\rms^2 2\rmp^5 3\rms 3\rmd  $ & $^2\rmD_{5/2} $ & 58.90430  & 59.29565   & 59.20474 & 59.28233 &  0.39135   & 0.30044  \\
 61  & $ 2\rms^2 2\rmp^5 3\rms 3\rmd  $ & $^2\rmF_{7/2} $ & 59.38727  & 59.35148   & 59.25137 & 59.33262 &  \-0.03579  & \-0.13590 \\
 62  & $   2\rms^2 2\rmp^5 3\rmp^2 $ & $^2\rmP_{1/2} $ &           & 59.41694   & 59.34593 & 59.41507 &            &          \\
 63  & $ 2\rms^2 2\rmp^5 3\rms 3\rmd  $ & $^2\rmD_{5/2} $ & 59.37816  & 59.42156   & 59.34082 & 59.42844 &  0.04340   & \-0.03734 \\
 64  & $   2\rms^2 2\rmp^5 3\rmp^2 $ & $^2\rmP_{1/2} $ &           & 59.52244   & 59.47971 & 59.55788 &            &          \\
 65  & $   2\rms^2 2\rmp^5 3\rmp^2 $ & $^2\rmP_{3/2} $ &           & 59.59573   & 59.55298 & 59.63194 &            &          \\
 66  & $ 2\rms^2 2\rmp^5 3\rms 3\rmd  $ & $^2\rmD_{3/2} $ &           & 59.73085   & 59.68822 & 59.74983 &            &          \\
 67  & $ 2\rms^2 2\rmp^5 3\rms 3\rmd  $ & $^2\rmP_{1/2} $ & 59.90670  & 59.95897   & 59.95160 & 60.01129 &  0.05227   & 0.04490  \\
 68  & $ 2\rms^2 2\rmp^5 3\rmp 3\rmd  $ & $^4\rmD_{1/2} $ &           & 60.13369   & 60.04611 & 60.13489 &            &          \\
 69  & $ 2\rms^2 2\rmp^5 3\rmp 3\rmd  $ & $^4\rmD_{3/2} $ &           & 60.20690   & 60.11829 & 60.19617 &            &          \\
 70  & $ 2\rms^2 2\rmp^5 3\rms 3\rmd  $ & $^2\rmF_{5/2} $ & 59.74267  & 60.22959   & 60.12842 & 60.20705 &  0.48692   & 0.38575  \\
 71  & $ 2\rms^2 2\rmp^5 3\rmp 3\rmd  $ & $^4\rmD_{5/2} $ &           & 60.32195   & 60.23238 & 60.32059 &            &          \\
 72  & $ 2\rms^2 2\rmp^5 3\rmp 3\rmd  $ & $^4\rmG_{7/2} $ &           & 60.43767   & 60.35233 & 60.43839 &            &          \\
 73  & $ 2\rms^2 2\rmp^5 3\rmp 3\rmd  $ & $^4\rmG_{9/2} $ &           & 60.46285   & 60.37302 & 60.46292 &            &          \\
 74  & $ 2\rms^2 2\rmp^5 3\rms 3\rmd  $ & $^2\rmP_{3/2} $ & 60.09806  & 60.46553   & 60.40230 & 60.47152 &  0.36747   & 0.30424  \\
 75  & $ 2\rms^2 2\rmp^5 3\rmp 3\rmd  $ & $^4\rmD_{7/2} $ &           & 60.47514   & 60.38502 & 60.47372 &            &          \\
 76  & $ 2\rms^2 2\rmp^5 3\rmp 3\rmd  $ & $^4\rmG_{11/2}$ &           & 60.50404   & 60.41011 & 60.50021 &            &          \\
 77  & $ 2\rms^2 2\rmp^5 3\rmp 3\rmd  $ & $^2\rmD_{5/2} $ &           & 60.52599   & 60.44777 & 60.53085 &            &          \\
 78  & $ 2\rms^2 2\rmp^5 3\rmp 3\rmd  $ & $^2\rmP_{3/2} $ &           & 60.57279   & 60.48675 & 60.57393 &            &          \\
 79  & $ 2\rms^2 2\rmp^5 3\rmp 3\rmd  $ & $^4\rmF_{5/2} $ &           & 60.63453   & 60.55605 & 60.63774 &            &          \\
 80  & $ 2\rms^2 2\rmp^5 3\rmp 3\rmd  $ & $^2\rmF_{7/2} $ &           & 60.63882   & 60.54820 & 60.64132 &            &          \\
 81  & $ 2\rms^2 2\rmp^5 3\rmp 3\rmd  $ & $^2\rmP_{1/2} $ &           & 60.69241   & 60.60183 & 60.69357 &            &          \\
 82  & $ 2\rms^2 2\rmp^5 3\rmp 3\rmd  $ & $^2\rmG_{7/2} $ &           & 60.74357   & 60.67077 & 60.75621 &            &          \\
 83  & $ 2\rms^2 2\rmp^5 3\rmp 3\rmd  $ & $^4\rmP_{1/2} $ &           & 60.83523   & 60.73533 & 60.82300 &            &          \\
 84  & $ 2\rms^2 2\rmp^5 3\rmp 3\rmd  $ & $^4\rmF_{9/2} $ &           & 60.84000   & 60.76830 & 60.85278 &            &          \\
 85  & $ 2\rms^2 2\rmp^5 3\rmp 3\rmd  $ & $^4\rmP_{3/2} $ &           & 60.86506   & 60.77171 & 60.85769 &            &          \\
 86  & $ 2\rms^2 2\rmp^5 3\rmp 3\rmd  $ & $^4\rmS_{3/2} $ &           & 60.90803   & 60.81615 & 60.90062 &            &          \\
 87  & $ 2\rms^2 2\rmp^5 3\rmp 3\rmd  $ & $^4\rmD_{7/2} $ &           & 60.92332   & 60.82874 & 60.91466 &            &          \\
 88  & $ 2\rms^2 2\rmp^5 3\rmp 3\rmd  $ & $^4\rmF_{5/2} $ &           & 60.94396   & 60.85344 & 60.93610 &            &          \\
 89  & $ 2\rms^2 2\rmp^5 3\rmp 3\rmd  $ & $^4\rmP_{5/2} $ &           & 60.95386   & 60.85914 & 60.94113 &            &          \\
 90  & $ 2\rms^2 2\rmp^5 3\rmp 3\rmd  $ & $^2\rmD_{3/2} $ &           & 60.98620   & 60.91494 & 60.99307 &            &          \\
 91  & $ 2\rms^2 2\rmp^5 3\rmp 3\rmd  $ & $^2\rmP_{3/2} $ &           & 61.05753   & 60.96448 & 61.04079 &            &          \\
 92  & $ 2\rms^2 2\rmp^5 3\rmp 3\rmd  $ & $^4\rmF_{9/2} $ &           & 61.06044   & 60.95573 & 61.04687 &            &          \\
 93  & $ 2\rms^2 2\rmp^5 3\rmp 3\rmd  $ & $^4\rmD_{5/2} $ &           & 61.09688   & 61.01801 & 61.09755 &            &          \\
 94  & $ 2\rms^2 2\rmp^5 3\rmp 3\rmd  $ & $^4\rmF_{7/2} $ &           & 61.12962   & 61.03764 & 61.12145 &            &          \\
\br
\end{tabular}
\end{indented}
\end{table}
\setcounter{table}{0}
\Table{-continued}
\br
Index & Configuration & $^{2S+1}L_J$ & NIST & AS & GRASP & FAC & AS--NIST & GRASP\\
& & & & & & & &--NIST \\
\mr
\095  & $ 2\rms^2 2\rmp^5 3\rmp 3\rmd  $ & $^2\rmF_{5/2} $ &           & 61.14734   & 61.06369 & 61.14624 &            &          \\
\096  & $ 2\rms^2 2\rmp^5 3\rmp 3\rmd  $ & $^4\rmD_{7/2} $ &           & 61.16998   & 61.08498 & 61.16228 &            &          \\
\097  & $ 2\rms^2 2\rmp^5 3\rmp 3\rmd  $ & $^2\rmP_{1/2} $ &           & 61.18777   & 61.11025 & 61.18841 &            &          \\
\098  & $ 2\rms^2 2\rmp^5 3\rmp 3\rmd  $ & $^2\rmD_{3/2} $ &           & 61.25489   & 61.18234 & 61.26136 &            &          \\
\099  & $ 2\rms^2 2\rmp^5 3\rmp 3\rmd  $ & $^4\rmD_{5/2} $ &           & 61.27670   & 61.20205 & 61.28070 &            &          \\
100  & $ 2\rms^2 2\rmp^5 3\rmp 3\rmd  $ & $^2\rmD_{3/2} $ &           & 61.32968   & 61.24893 & 61.32473 &            &          \\
101  & $ 2\rms^2 2\rmp^5 3\rmp 3\rmd  $ & $^4\rmG_{5/2} $ &           & 61.34088   & 61.24585 & 61.33104 &            &          \\
102  & $ 2\rms^2 2\rmp^5 3\rmp 3\rmd  $ & $^4\rmD_{1/2} $ &           & 61.41249   & 61.35321 & 61.42349 &            &          \\
103  & $ 2\rms^2 2\rmp^5 3\rmp 3\rmd  $ & $^2\rmF_{5/2} $ &           & 61.43393   & 61.34445 & 61.42516 &            &          \\
104  & $ 2\rms^2 2\rmp^5 3\rmp 3\rmd  $ & $^2\rmS_{1/2} $ &           & 61.49422   & 61.42443 & 61.48894 &            &          \\
105  & $ 2\rms^2 2\rmp^5 3\rmp 3\rmd  $ & $^4\rmF_{3/2} $ &           & 61.49958   & 61.40208 & 61.50040 &            &          \\
106  & $ 2\rms^2 2\rmp^5 3\rmp 3\rmd  $ & $^4\rmF_{7/2} $ &           & 61.51701   & 61.42221 & 61.50676 &            &          \\
107  & $ 2\rms^2 2\rmp^5 3\rmp 3\rmd  $ & $^2\rmF_{5/2} $ &           & 61.57211   & 61.47347 & 61.55745 &            &          \\
108  & $ 2\rms^2 2\rmp^5 3\rmp 3\rmd  $ & $^2\rmF_{7/2} $ &           & 61.65079   & 61.56056 & 61.64055 &            &          \\
109  & $ 2\rms^2 2\rmp^5 3\rmp 3\rmd  $ & $^2\rmG_{9/2} $ &           & 61.67835   & 61.58175 & 61.66370 &            &          \\
110  & $ 2\rms^2 2\rmp^5 3\rmp 3\rmd  $ & $^4\rmD_{3/2} $ &           & 61.69701   & 61.61493 & 61.68724 &            &          \\
111  & $ 2\rms^2 2\rmp^5 3\rmp 3\rmd  $ & $^2\rmG_{9/2} $ &           & 61.79328   & 61.69334 & 61.76535 &            &          \\
112  & $ 2\rms^2 2\rmp^5 3\rmp 3\rmd  $ & $^4\rmF_{3/2} $ &           & 61.82696   & 61.72111 & 61.80285 &            &          \\
113  & $ 2\rms^2 2\rmp^5 3\rmp 3\rmd  $ & $^2\rmD_{5/2} $ &           & 61.89847   & 61.80037 & 61.87459 &            &          \\
114  & $ 2\rms^2 2\rmp^5 3\rmp 3\rmd  $ & $^2\rmD_{5/2} $ &           & 61.96161   & 61.86516 & 61.93945 &            &          \\
115  & $ 2\rms^2 2\rmp^5 3\rmp 3\rmd  $ & $^4\rmP_{1/2} $ &           & 61.97606   & 61.88979 & 61.96275 &            &          \\
116  & $ 2\rms^2 2\rmp^5 3\rmp 3\rmd  $ & $^2\rmF_{7/2} $ &           & 61.99048   & 61.89980 & 61.96894 &            &          \\
117  & $ 2\rms^2 2\rmp^5 3\rmp 3\rmd  $ & $^2\rmP_{3/2} $ &           & 62.00886   & 61.91457 & 61.98790 &            &          \\
118  & $ 2\rms^2 2\rmp^5 3\rmp 3\rmd  $ & $^4\rmP_{5/2} $ &           & 62.04487   & 61.94888 & 62.02041 &            &          \\
119  & $ 2\rms^2 2\rmp^5 3\rmp 3\rmd  $ & $^4\rmD_{1/2} $ &           & 62.06520   & 61.97998 & 62.05719 &            &          \\
120  & $ 2\rms^2 2\rmp^5 3\rmp 3\rmd  $ & $^4\rmD_{3/2} $ &           & 62.06835   & 61.97802 & 62.06089 &            &          \\
121  & $ 2\rms^2 2\rmp^5 3\rmp 3\rmd  $ & $^2\rmF_{7/2} $ &           & 62.10521   & 61.99687 & 62.07502 &            &          \\
122  & $ 2\rms^2 2\rmp^5 3\rmp 3\rmd  $ & $^2\rmD_{5/2} $ &           & 62.11542   & 62.02163 & 62.09605 &            &          \\
123  & $ 2\rms^2 2\rmp^5 3\rmp 3\rmd  $ & $^2\rmD_{3/2} $ &           & 62.12317   & 62.05018 & 62.11336 &            &          \\
124  & $ 2\rms^2 2\rmp^5 3\rmp 3\rmd  $ & $^2\rmD_{3/2} $ &           & 62.29418   & 62.27680 & 62.32294 &            &          \\
125  & $ 2\rms^2 2\rmp^5 3\rmp 3\rmd  $ & $^2\rmD_{5/2} $ &           & 62.33871   & 62.28342 & 62.34267 &            &          \\
126  & $ 2\rms^2 2\rmp^5 3\rmp 3\rmd  $ & $^2\rmP_{1/2} $ &           & 62.36979   & 62.31859 & 62.37920 &            &          \\
127  & $ 2\rms^2 2\rmp^5 3\rmp 3\rmd  $ & $^2\rmP_{3/2} $ &           & 62.54002   & 62.50435 & 62.55742 &            &          \\
128  & $ 2\rms^2 2\rmp^5 3\rmp 3\rmd  $ & $^2\rmS_{1/2} $ &           & 62.71307   & 62.70084 & 62.74891 &            &          \\
129  & $ 2\rms^2 2\rmp^5 3\rmp 3\rmd  $ & $^2\rmG_{7/2} $ &           & 62.83419   & 62.72376 & 62.79211 &            &          \\
130  & $ 2\rms^2 2\rmp^5 3\rmp 3\rmd  $ & $^2\rmF_{5/2} $ &           & 62.89339   & 62.83399 & 62.88208 &            &          \\
131  & $ 2\rms^2 2\rmp^5 3\rmp 3\rmd  $ & $^4\rmP_{3/2} $ &           & 62.94192   & 62.86270 & 62.92242 &            &          \\
132  & $ 2\rms^2 2\rmp^5 3\rmp 3\rmd  $ & $^2\rmP_{1/2} $ &           & 63.10850   & 63.03638 & 63.08932 &            &          \\
133  & $ 2\rms^2 2\rmp^5 3\rmp 3\rmd  $ & $^2\rmD_{3/2} $ &           & 63.27928   & 63.21048 & 63.29528 &            &          \\
134  & $ 2\rms^2 2\rmp^5 3\rmp 3\rmd  $ & $^2\rmD_{5/2} $ &           & 63.35811   & 63.27358 & 63.36869 &            &          \\
\br
\end{tabular}
 \item[]$^{\rma}${\sf http://physics.nist.gov/PhysRefData/ASD/levels\_form.html}
 \item[]$^{\rmb}${\sc autostructure} (present work).
 \item[]$^{\rmc}$Aggarwal and Keenan (2007).
 \item[]$^{\rmd}$Present work.
\end{indented}
\end{table}
\clearpage

\Table{\label{a-coeff}Electric dipole radiative rates.}  
\br
 $i$ & $j$ & AS$^\rma$  & GRASP$^\rmb$ & FAC$^\rmc$ \\
\mr
  1 &   2  & 6.071(09)$^\rmd$ & 6.283(09)  &  6.303(09)  \\
  1 &   3  & 7.542(09) & 7.834(09)  &  7.884(09)  \\
  1 &   6  & 8.553(11) & 8.202(11)  &  6.465(11)  \\
  1 &   7  & 8.317(11) & 8.450(11)  &  6.666(11)  \\
  1 &  26  & 9.461(10) & 9.327(10)  &  8.946(10)  \\
  1 &  27  & 3.088(11) & 3.087(11)  &  2.991(11)  \\
  1 &  30  & 1.199(11) & 1.187(11)  &  1.188(11)  \\
  1 &  33  & 8.400(10) & 8.529(10)  &  8.788(10)  \\
  1 &  36  & 6.020(10) & 5.136(10)  &  6.060(10)  \\
  1 &  37  & 2.067(10) & 1.664(10)  &  2.180(10)  \\
  1 &  38  & 4.509(10) & 4.774(10)  &  4.486(10)  \\
  1 &  39  & 9.727(10) & 9.867(10)  &  9.301(10)  \\
  1 &  41  & 8.636(09) & 6.427(09)  &  7.977(09)  \\
  1 &  47  & 2.700(10) & 4.872(09)  &  5.000(10)  \\
  1 &  48  & 3.379(10) & 4.862(10)  &  5.315(09)  \\
  1 &  51  & 1.473(10) & 5.425(08)  &  1.757(08)  \\
  1 &  53  & 1.161(12) & 1.025(12)  &  1.000(12)  \\
  1 &  54  & 4.042(12) & 3.670(12)  &  3.545(12)  \\
  1 &  55  & 3.413(12) & 3.161(12)  &  3.205(12)  \\
  1 &  56  & 1.197(12) & 7.926(11)  &  9.465(11)  \\
  1 &  57  & 2.133(11) & 2.632(11)  &  2.790(11)  \\
  1 &  62  & 3.950(11) & 5.031(11)  &  4.780(11)  \\
  1 &  64  & 7.181(10) & 4.447(10)  &  1.723(11)  \\
  1 &  65  & 2.730(12) & 2.170(12)  &  3.221(12)  \\
  1 &  66  & 1.217(13) & 1.359(13)  &  1.145(13)  \\
  1 &  67  & 2.254(13) & 2.461(13)  &  2.260(13)  \\
  1 &  74  & 7.338(12) & 9.012(12)  &  8.160(12)  \\
\br
\end{tabular}
 \item[]$^{\rma}${\sc autostructure} (present work).
 \item[]$^{\rmb}$Aggarwal and Keenan (2007).
 \item[]$^{\rmc}$Present work.
 \item[]$^{\rmd}$(m) denotes $\times 10^m$.
\end{indented}
\end{table}

\Table{\label{diff_tran}Excitation energies (Ryd) used for dipole transitions\\
 with large  differences at low temperatures between \\
the ICFT and DARC effective collision strengths.}
\br
 $i$ & $j$ & ICFT$^\rma$ & DARC$^\rmb$ & ICFT--DARC \\
\mr
   1 &  6  &  52.60745 &  52.31155 &  0.29589  \\
   1 &  7  &  53.51871 &  53.22823 &  0.29048  \\
   2 &  8  &  51.74576 &  51.51313 &  0.23263  \\
   2 & 11  &  52.05286 &  51.83545 &  0.21741  \\
   2 & 12  &  52.15916 &  51.95366 &  0.20550  \\
   2 & 15  &  52.53278 &  52.28154 &  0.25124  \\
   2 & 17  &  52.96107 &  52.73537 &  0.22570  \\
   2 & 22  &  53.32558 &  53.01148 &  0.31410  \\
   2 & 23  &  53.92686 &  53.69872 &  0.22814  \\
   2 & 24  &  54.12749 &  53.85740 &  0.27009  \\
   3 &  8  &  51.55486 &  51.33900 &  0.21586  \\
   3 &  9  &  51.79511 &  51.57144 &  0.22367  \\
   3 & 11  &  51.86196 &  51.62987 &  0.23209  \\
   3 & 13  &  52.08980 &  51.86496 &  0.22485  \\
   3 & 14  &  52.07754 &  51.86496 &  0.21259  \\
   3 & 15  &  52.34188 &  52.10219 &  0.23969  \\
   3 & 17  &  52.77017 &  52.55290 &  0.21727  \\
   3 & 19  &  52.91628 &  52.67441 &  0.24187  \\
   3 & 22  &  53.13468 &  52.82709 &  0.30759  \\
   3 & 23  &  53.73596 &  53.50953 &  0.22643  \\
   3 & 24  &  53.93659 &  53.66709 &  0.26949  \\
   4 &  6  &  46.45183 &  46.11676 &  0.33507  \\
   4 &  7  &  47.36309 &  47.04529 &  0.31780  \\
   5 &  6  &  46.42536 &  46.09344 &  0.33192  \\
\br
\end{tabular}
 \item[]$^{\rma}$Present work.
 \item[]$^{\rmb}$Aggarwal and Keenan (2008).
\end{indented}
\end{table}
\clearpage

\Table{\label{upsdamp}Undamped (U) and Auger-plus-radiation damped (A+R) \\
effective collision strengths
$\Upsilon_{i,j}$, at the given temperatures.}
\br
 & & \multicolumn{2}{c}{5.12$\times10^4$~K} &
 \multicolumn{2}{c}{5.12$\times10^5$~K} &
 \multicolumn{2}{c}{5.12$\times10^6$~K} \\
 \cline{3-4}  \cline{5-6}  \cline{7-8} \\
$i$ & $j$ & U & A+R & U & A+R & U & A+R \\
\mr
1  &\02  & 1.38(+0)$^\rma$& 1.37(+0) & 1.22(+0) & 1.22(+0) & 1.51(+0) & 1.51(+0)  \\
1  &\03  & 2.28(+0) & 2.30(+0) & 2.40(+0) & 2.40(+0) & 2.99(+0) & 2.99(+0)  \\
1  &\04  & 1.27($-$1) & 1.27($-$1) & 1.28($-$1) & 1.28($-$1) & 1.46($-$1) & 1.43($-$1)  \\
1  &\05  & 1.90($-$1) & 1.90($-$1) & 1.92($-$1) & 1.92($-$1) & 2.19($-$1) & 2.14($-$1)  \\
1  &\06  & 1.56($-$1) & 5.25($-$2) & 8.28($-$2) & 3.08($-$2) & 1.64($-$2) & 9.04($-$3)  \\
1  &\07  & 8.20($-$2) & 3.67($-$2) & 4.12($-$2) & 1.79($-$2) & 8.38($-$3) & 4.87($-$3)  \\
1  &\08  & 5.58($-$2) & 3.08($-$2) & 2.17($-$2) & 1.43($-$2) & 6.27($-$3) & 5.29($-$3)  \\
1  &\09  & 5.32($-$2) & 2.88($-$2) & 2.06($-$2) & 1.21($-$2) & 5.56($-$3) & 4.39($-$3)  \\
1  & 10  & 5.39($-$2) & 3.69($-$2) & 2.12($-$2) & 1.54($-$2) & 6.28($-$3) & 5.48($-$3)  \\
1  & 11  & 3.14($-$2) & 1.67($-$2) & 1.34($-$2) & 7.34($-$3) & 4.12($-$3) & 3.26($-$3)  \\
1  & 12  & 1.74($-$2) & 7.07($-$3) & 7.60($-$3) & 3.75($-$3) & 2.13($-$3) & 1.60($-$3)  \\
1  & 13  & 3.84($-$2) & 1.47($-$2) & 1.45($-$2) & 7.08($-$3) & 4.02($-$3) & 3.03($-$3)  \\
1  & 14  & 4.00($-$2) & 1.85($-$2) & 1.67($-$2) & 9.85($-$3) & 4.99($-$3) & 4.08($-$3)  \\
1  & 15  & 4.19($-$2) & 2.27($-$2) & 1.75($-$2) & 1.12($-$2) & 7.94($-$3) & 7.04($-$3)  \\
1  & 16  & 1.38($-$2) & 7.14($-$3) & 6.78($-$3) & 3.60($-$3) & 1.73($-$3) & 1.25($-$3)  \\
1  & 17  & 2.66($-$2) & 1.39($-$2) & 1.32($-$2) & 6.97($-$3) & 3.73($-$3) & 2.82($-$3)  \\
1  & 18  & 1.51($-$2) & 6.32($-$3) & 7.76($-$3) & 3.53($-$3) & 2.21($-$3) & 1.60($-$3)  \\
1  & 20  & 2.67($-$2) & 1.39($-$2) & 1.46($-$2) & 7.66($-$3) & 3.66($-$3) & 2.61($-$3)  \\
1  & 19  & 3.71($-$2) & 2.09($-$2) & 1.70($-$2) & 1.02($-$2) & 5.99($-$3) & 5.00($-$3)  \\
1  & 21  & 4.38($-$2) & 2.24($-$2) & 1.94($-$2) & 9.99($-$3) & 5.69($-$3) & 4.34($-$3)  \\
1  & 22  & 3.45($-$2) & 1.46($-$2) & 1.59($-$2) & 7.09($-$3) & 3.53($-$3) & 2.26($-$3)  \\
1  & 23  & 3.86($-$2) & 2.63($-$2) & 3.22($-$2) & 2.30($-$2) & 1.90($-$2) & 1.75($-$2)  \\
1  & 24  & 2.42($-$2) & 1.10($-$2) & 1.02($-$2) & 5.86($-$3) & 3.22($-$3) & 2.62($-$3)  \\
1  & 38  & 7.23($-$3) & 3.76($-$3) & 2.95($-$3) & 1.83($-$3) & 1.41($-$3) & 1.27($-$3)  \\
1  & 39  & 4.26($-$3) & 2.46($-$3) & 2.36($-$3) & 1.76($-$3) & 1.73($-$3) & 1.65($-$3)  \\
1  & 25  & 9.07($-$2) & 8.37($-$2) & 8.61($-$2) & 8.08($-$2) & 8.35($-$2) & 8.26($-$2)  \\
1  & 43  & 7.58($-$3) & 4.53($-$3) & 2.15($-$3) & 1.13($-$3) & 2.98($-$4) & 1.71($-$4)  \\
1  & 42  & 6.52($-$3) & 3.70($-$3) & 3.23($-$3) & 2.05($-$3) & 1.23($-$3) & 1.08($-$3)  \\
1  & 44  & 7.51($-$3) & 3.65($-$3) & 3.01($-$3) & 1.79($-$3) & 1.44($-$3) & 1.30($-$3)  \\
1  & 46  & 6.50($-$3) & 3.02($-$3) & 2.49($-$3) & 1.31($-$3) & 6.90($-$4) & 5.45($-$4)  \\
1  & 48  & 5.04($-$3) & 1.88($-$3) & 1.35($-$3) & 4.82($-$4) & 1.84($-$4) & 7.99($-$5)  \\
1  & 53  & 3.10($-$3) & 1.27($-$3) & 1.13($-$3) & 6.25($-$4) & 5.14($-$4) & 4.53($-$4)  \\
1  & 49  & 5.05($-$3) & 2.67($-$3) & 1.65($-$3) & 8.32($-$4) & 3.54($-$4) & 2.52($-$4)  \\
1  & 52  & 5.34($-$3) & 2.47($-$3) & 1.70($-$3) & 7.64($-$4) & 3.08($-$4) & 1.93($-$4)  \\
1  & 54  & 3.65($-$3) & 1.47($-$3) & 1.21($-$3) & 6.31($-$4) & 4.50($-$4) & 3.82($-$4)  \\
1  & 58  & 5.18($-$3) & 2.01($-$3) & 1.81($-$3) & 9.03($-$4) & 5.36($-$4) & 4.29($-$4)  \\
1  & 55  & 6.96($-$3) & 5.05($-$3) & 4.61($-$3) & 4.11($-$3) & 2.80($-$3) & 2.74($-$3)  \\
1  & 59  & 1.17($-$2) & 9.16($-$3) & 8.58($-$3) & 7.80($-$3) & 5.53($-$3) & 5.44($-$3)  \\
1  & 56  & 5.52($-$3) & 2.43($-$3) & 2.39($-$3) & 1.38($-$3) & 8.74($-$4) & 7.50($-$4)  \\
1  & 57  & 4.91($-$3) & 1.66($-$3) & 1.94($-$3) & 8.33($-$4) & 5.58($-$4) & 4.21($-$4)  \\
1  & 60  & 1.27($-$2) & 9.56($-$3) & 9.18($-$3) & 8.14($-$3) & 5.29($-$3) & 5.16($-$3)  \\
1  & 63  & 1.31($-$2) & 1.04($-$2) & 9.78($-$3) & 8.92($-$3) & 5.89($-$3) & 5.79($-$3)  \\
1  & 61  & 1.01($-$2) & 7.06($-$3) & 6.92($-$3) & 6.00($-$3) & 4.13($-$3) & 4.02($-$3)  \\
1  & 66  & 1.65($-$3) & 5.00($-$4) & 5.56($-$4) & 1.51($-$4) & 7.43($-$5) & 2.45($-$5)  \\
1  & 70  & 9.28($-$3) & 5.77($-$3) & 5.58($-$3) & 4.62($-$3) & 3.36($-$3) & 3.25($-$3)  \\
1  & 67  & 5.48($-$3) & 1.67($-$3) & 1.82($-$3) & 8.44($-$4) & 5.88($-$4) & 4.73($-$4)  \\
1  & 74  & 5.77($-$3) & 3.10($-$3) & 3.20($-$3) & 2.45($-$3) & 1.71($-$3) & 1.62($-$3)  \\
\br
\end{tabular}
 \item[]$^{\rma}$(m) denotes $\times 10^m$.
\end{indented}
\end{table}
\clearpage

\clearpage

\end{document}